\documentclass[sigconf, final]{acmart}
\AtBeginDocument{%
  }

\copyrightyear{2026}
\acmYear{2026}
\setcopyright{cc}
\setcctype{by}
\acmConference[CHI '26]{Proceedings of the 2026 CHI Conference on Human Factors in Computing Systems}{April 13--17, 2026}{Barcelona, Spain}
\acmBooktitle{Proceedings of the 2026 CHI Conference on Human Factors in Computing Systems (CHI '26), April 13--17, 2026, Barcelona, Spain}
\acmPrice{}
\acmDOI{10.1145/3772318.3790388}
\acmISBN{979-8-4007-2278-3/2026/04}

\usepackage{framed}

\begin{document}

\title{Vibe Check: Understanding the Effects of LLM-Based Conversational Agents' Personality and Alignment  on User Perceptions in Goal-Oriented Tasks}

\author{Hasibur Rahman}
\affiliation{%
  \institution{Northeastern University}
  \city{Boston}
  \state{Massachusetts}
  \country{USA}}
  \email{rahman.has@northeastern.edu}

\author{Smit Desai}
\authornote{Corresponding author}
\affiliation{
  \institution{Northeastern University}
  \city{Boston}
  \state{Massachusetts}
  \country{USA}}
\email{sm.desai@northeastern.edu}

\renewcommand{\shorttitle}{Effects of LLM-based CAs' Personality on User Perceptions}

\begin{abstract}

Large language models (LLMs) enable conversational agents (CAs) to express distinctive personalities, raising new questions about how such designs shape user perceptions. This study investigates how personality expression levels and user-agent personality alignment influence perceptions in goal-oriented tasks. In a between-subjects experiment (N=150), participants completed travel planning with CAs exhibiting low, medium, or high expression across the Big Five traits, controlled via our novel Trait Modulation Keys framework. Results revealed an inverted-U relationship: medium expression produced the most positive evaluations across Intelligence, Enjoyment, Anthropomorphism, Intention to Adopt, Trust, and Likeability, significantly outperforming both extremes. Personality alignment further enhanced outcomes, with Extraversion and Emotional Stability emerging as the most influential traits. Cluster analysis identified three distinct compatibility profiles, with "Well-Aligned" users reporting substantially positive perceptions. These findings demonstrate that personality expression and strategic trait alignment constitute optimal design targets for CA personality, offering design implications as LLM-based CAs become increasingly prevalent.

\end{abstract}



\begin{teaserfigure}
  \includegraphics[height=330pt, width=\textwidth]{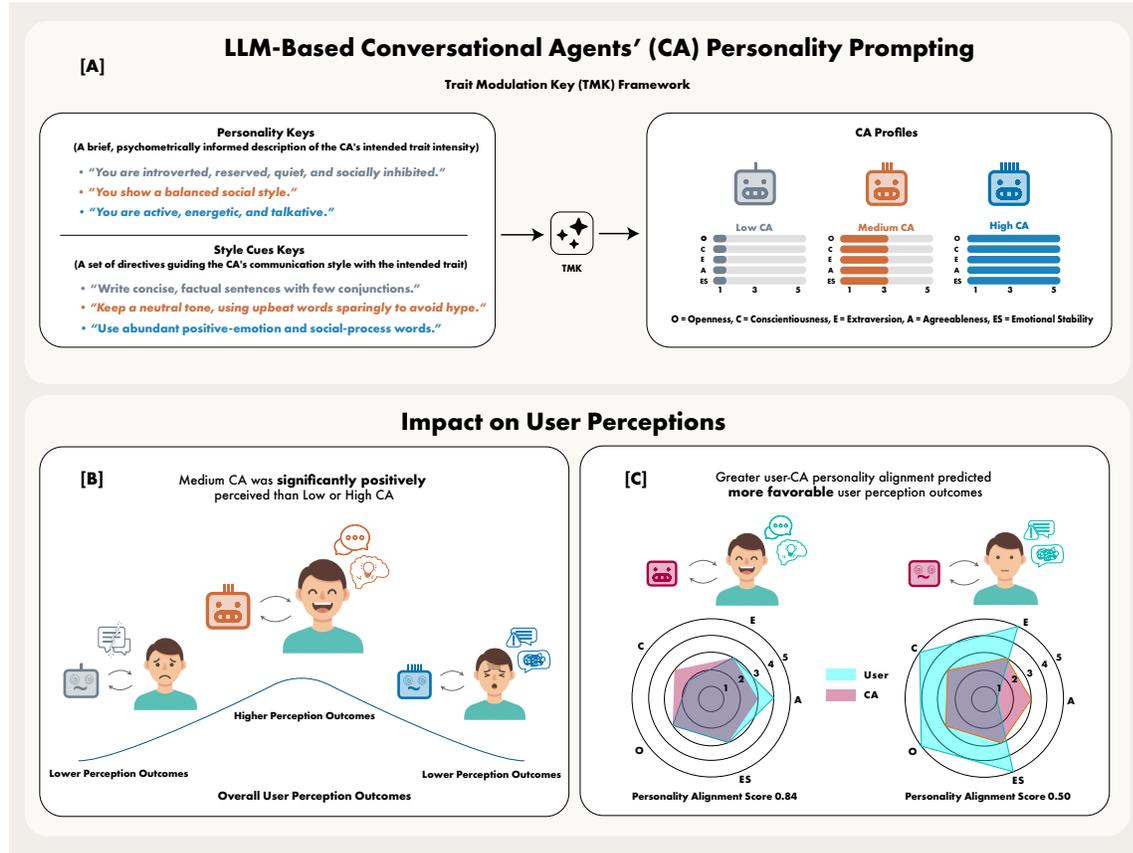}
  \caption{Overview of our Large Language Model (LLM)-based Conversational Agents' (CA) personality prompting and its impact on user perceptions. [A] Our novel Trait Modulation Keys (TMK) prompting framework pairs a Personality Key with Style Cues Keys to steer Big-Five trait intensities to design three specific CAs (Low/Medium/High). [B] In a goal-oriented travel-planning study (N=150), the Medium CA yielded the most favorable perception outcomes across perceived Intelligence, Enjoyment, Anthropomorphism, Intention to Adopt, Trust, and Likeability.
  [C] Greater user–CA personality alignment further improved perceptions, illustrated via higher alignment scores.}
  \Description{Infographic with two rows summarizing the study design and findings. Top row—“LLM-Based CA Personality Prompting”: A left card lists example Personality Keys (e.g., “introverted…,” “balanced…,” “energetic…talkative”) and Style Cues Keys (e.g., concise, neutral tone, limited hype, positive social wording). An arrow labeled TMK points to a right card titled “CA Conditions,” showing three CA profiles (robot icons) with sliders for the Big Five (O, C, E, A, ES; scale 1–5). There are three CA profiles: Low condition, Medium condition, and High condition. For Low condition CA all O, C, E, A, ES are set at 1; for Medium condition CA all O, C, E, A, ES are set at 3; and for High condition CA all O, C, E, A, ES are set at 5. Bottom row—“Impact on User Perceptions”: Left panel depicts that the Medium condition CA is perceived most positively in a goal-oriented travel-planning study (n=150), with an inverted-U schematic and emoji/user icons; outcomes noted include Intelligence, Enjoyment, Anthropomorphism, Intention to Adopt, Trust, and Likeability. Right panel illustrates alignment effects: two radar charts overlay User vs. CA personality traits. The higher-overlap chart (alignment score =0.84) sits beside a smiling user; the lower-overlap chart (=0.50) beside a neutral user—conveying that greater user–CA personality alignment yields more favorable perceptions.}
  \label{fig:teaser}
\end{teaserfigure}

\maketitle

\section{Introduction}

In 1966, Joseph Weizenbaum introduced \textit{ELIZA}, one of the earliest conversational agents (CAs). ELIZA simulated a Rogerian psychotherapist by reflecting user statements back in the form of questions. The system contained no understanding of language or intelligence—only simple pattern-matching and string substitution rules \cite{weizenbaum_elizacomputer_1966}. Despite its simplicity, users often treated ELIZA as more than a machine. Weizenbaum reported that his secretary, fully aware of the system’s limitations, nonetheless asked to be left alone with it to continue her conversation. This episode revealed a persistent feature of human–computer interaction: when computers engage us in dialogue, we readily perceive personality and social presence, even in artificial systems. 

The Computers Are Social Actors (CASA) paradigm later formalized this phenomenon. Through controlled experiments, \citet{nass_computers_1994} demonstrated that people automatically apply social rules and trait attributions to interactive systems—showing politeness, reciprocating self-disclosure, and expressing preference for systems that mirror their own personalities. This effect not only persists but intensifies in interactions with CAs \cite{chin_being_2021, xu_deep_2022, heyselaar_casa_2023}. Interestingly, such responses occur automatically, even among experienced users who explicitly deny anthropomorphizing machines \cite{nass_computers_1994, desai_metaphors_2023, pradhan_phantom_2019}. The implication is clear: if people treat computers as social actors, then the personality that CAs project can also systematically shape user perceptions.

Today, millions of people interact with large language model (LLM)-based CAs such as ChatGPT, Claude, and Gemini for tasks ranging from travel planning to creative writing \cite{report2025}. During these interactions, users encounter distinctive communication styles that they interpret as personality. Although companies invest heavily in making these agents ``helpful,'' ``harmless,'' and ``friendly,'' \cite{yuan_hard_2025, bai_training_2022}, we lack a systematic understanding of how personality expression actually shapes user perceptions. This deficit becomes increasingly consequential as CAs transition from optional utilities to critical infrastructure in education, healthcare, and professional practice. Unlike ELIZA’s scripted responses, today’s LLM-based agents represent a fundamental shift: personality is no longer an incidental byproduct of surface wording but can be deliberately designed and systematically controlled through natural language prompting \cite{ramirez_controlling_2023, jiang_evaluating_2023, jiang_personallm_2024, serapio-garcia_personality_2025}. This technological advance creates both new opportunities and pressing questions: how should personality be expressed in CAs, at what intensity, and in what relation to the user’s own traits?  

Prior HCI research has shown that agent personality designed with linguistic cues influences user perceptions \cite{zhou_trusting_2019, volkel_user_2022, volkel_examining_2021, shumanov_making_2021, ruane_user_2020}, typically operationalized through the Big Five Personality Traits (Openness, Conscientiousness, Extraversion, Agreeableness, Neuroticism/Emotional Stability) \cite{john_big_1991, john_big-five_1999}. However, this foundational work relied on tightly scripted dialogues that manipulated traits at bipolar extremes (e.g., agreeable/disagreeable \cite{volkel_examining_2021}) \cite{volkel_user_2022, ruane_user_2020}. While these designs afforded experimental precision, they leave a critical blind spot: we know little about how users respond to moderate or balanced agent personalities, as pointed out by \citet{volkel_examining_2021}. 

Research on personality alignment—the compatibility between user and agent personality traits—has attempted to extend this work. The widely purported "similarity-attraction" hypothesis suggests that users should prefer agents matching their own personalities \cite{moon_how_1996, nass_does_2001, braun_at_2019}, and some evidence supports enhanced engagement when agent extraversion mirrors the user’s \cite{gnewuch_effect_2020, shumanov_making_2021, braun_at_2019}. Yet findings remain mixed: some studies report no benefit or even preferences for complementary personalities \cite{isbister_consistency_2000, spagnolli_similarity_2025}, while others suggest that alignment effects vary by trait and context \cite{volkel_examining_2021, amin_kuhail_assessing_2024}. The mixed evidence may reflect how alignment has been operationalized: prior work typically employs binary matching/mismatching treatments rather than recognizing compatibility as a continuous, multidimensional phenomenon. Addressing this limitation opens the door to more precise explanations of personality alignment effects and their contextual variation.

LLMs expand what is practically possible for personality-aware CA design through prompt-based steering. Existing studies demonstrate that prompts can reliably induce Big Five traits \cite{ramirez_controlling_2023, jiang_evaluating_2023} and even coordinate multiple traits concurrently \cite{jiang_personallm_2024, serapio-garcia_personality_2025}. However, most approaches remain limited: many rely on persona-based biographical backstories that undermine role fidelity \cite{serapio-garcia_personality_2025}, and most concentrate control at binary endpoints (high or low) \cite{jiang_evaluating_2023, jiang_personallm_2024}. The capacity for nuanced, medium-level personality expression has not been systematically examined.  

We address these gaps through a controlled, between-subjects study of personality expression levels and personality alignment effects in goal-oriented CAs. To this end, we develop \textit{Trait Modulation Keys} (TMK), a modular prompting framework that simultaneously controls all Big Five traits at low, medium, and high expression levels without requiring biographical backstories. TMK enables the first systematic comparison of personality expression levels beyond the binary extremes and reliance on persona-based backstories that have constrained prior research. An overview of our framework and main findings is shown in Figure \ref{fig:teaser}. 

We implement a 3×1 between-subjects design (N=150, $n=50$ per condition) in which participants interact with a travel planning assistant configured with one of three personality profiles: all traits set to low, medium, or high expression. This resulted in three CAs, which we refer to as Low CA, Medium CA, and High CA, each exhibiting distinct personality expression levels. We focus on goal-oriented tasks because such contexts provide concrete subtasks and success criteria, allowing fair cross-condition comparisons while minimizing confounds from interface affordances \cite{atzmuller_experimental_2010, haresamudram_tasks_2025, nguyen_user_2022}. After the interaction, participants rated six user perception measures—Intelligence, Enjoyment, Anthropomorphism, Intention to Adopt, Trust, and Likeability—selected based on \citet{10.1145/3571884.3597139}'s framework linking user perceptions to agent relationships. In addition, we measured each participant’s own Big Five traits to calculate personality alignment, operationalized as the normalized Euclidean distance across all five traits between user and agent personalities.

This study design allows us to move beyond prior binary manipulations of personality expression and categorical treatments of personality alignment to examine both in greater depth. Guided by these gaps, our study examines four research questions (RQs): \textbf{RQ1:} \textit{How do different CA personality expressions affect user perceptions in goal-oriented tasks?} \textbf{RQ2:} \textit{To what extent does personality alignment between users and CAs influence user perceptions?} \textbf{RQ3:} \textit{How do specific personality trait alignments between users and CAs contribute to user perceptions?} and \textbf{RQ4:} \textit{Do users naturally form distinct groups based on their personality alignment patterns with CAs, and how do these groups differ in their perceptions?} Together, these questions enable a comprehensive investigation that progresses from general effects of personality expression, to mechanisms of compatibility, to user-level differences with design implications.

 Our results highlight systematic effects of medium expression levels, trait-specific patterns of alignment, and emergent user profiles. Building on these findings, our work contributes the following:
\begin{itemize}
    \item \textbf{A Modular Prompting Framework for LLM Personality Expression.} We present TMK, a systematic method for generating consistent personality profiles in CAs, including medium personality that addresses gaps in existing high/low manipulation methods.  
    \item \textbf{Empirical Evidence for Personality Effects in Goal-Oriented CA Interaction.} Through a controlled study, we show that CA personalities systematically influence user perceptions. Specifically, Medium CA consistently produced significantly higher perceived Intelligence, Enjoyment, Anthropomorphism, Intention to Adopt, Trust, and Likeability than Low CA, and also yielded significantly greater perceived Intelligence and Likeability than High CA. This pattern reveals an inverted-U relationship between personality expression level and user experience, with the medium level of personality expression outperforming extremes.
    \item \textbf{Personality Alignment as a Design Consideration.} We establish personality alignment as a key design factor, showing consistent positive associations with user perceptions. Individual traits showed differential effects, with misalignment in Extraversion and Conscientiousness most broadly and negatively affecting user perceptions, while Openness misalignment had comparatively minimal effect.
 
\end{itemize}

\section{Related Work}

Our related work examines four domains essential to understanding personality effects in CA design: foundational research on CA personality expression and user perceptions, personality alignment between users and agents, advances in LLM-based personality prompting, and goal-oriented conversational contexts. Together, these domains establish the theoretical foundation and practical capabilities needed to investigate how CA personality and alignment jointly influence user perceptions in goal-oriented tasks. 

\subsection{Foundations of CA Personality Research in HCI}

People naturally perceive and ascribe personality traits to CAs, even when designers do not explicitly specify them \cite{Kuzminykh_Sun_Govindaraju_Avery_Lank_2020}. This tendency stems from the CASA paradigm, which demonstrates that users apply social rules and personality perceptions to machines in much the same way as they do to humans \cite{nass_computers_1994}. Early work by \citet{nass_does_2001} established that an agent's demeanor shapes user responses: extroverted voice interfaces were preferred by extroverted users in one experiment. However, with text-based interfaces, the effect was less pronounced, revealing that conveying personality effectively through language alone requires careful design \cite{nass_does_2001}. This design challenge has motivated extensive HCI research spanning decades, investigating how to implement agent personalities and their impact on user perceptions.

The Big Five Personality Traits\footnote{also referred to by the acronym OCEAN}—Openness (curiosity and creativity), Conscientiousness (organization and diligence), Extraversion (sociability and energy), Agreeableness (cooperation and wa\-rmth), and Neuroticism (anxiety and distress, the opposite of Emotional Stability) \cite{john_big_1991, john_big-five_1999}—have become the dominant framework for defining and manipulating such personalities in HCI literature \cite{isbister_consistency_2000, zhou_trusting_2019, volkel_user_2022, spagnolli_similarity_2025}. Beyond CAs, Big Five traits and their sub-facets have also been used to characterize behavior in robots and embodied agents, with evidence that trait expressions are relatively stable across cultures \cite{nass_wired_2005, schmitz_modelling_2007, trouvain_modelling_2006, vinciarelli_survey_2014, deary_trait_2009, costa_four_1992}. From a psycholinguistic perspective, personality is inferred from consistent verbal and non-verbal markers \cite{scherer_personality_1979, argyle_bodily_1988}; extensive work links personality traits to language features across essays, narratives, and text messages using correlational and group-comparison designs \cite{gill_taking_2002, hirsh_personality_2009, holtgraves_text_2011, boyd_language-based_2017, carment_persuasiveness_1965, dewaele_personality_2000, furnham_language_1990, mehl_personality_2006, oberlander_individual_2004, patterson_social_1966, pennebaker_linguistic_1999}.

Thus far, most CA implementations of the Big Five Traits operationalize personality with tightly controlled, pre-scripted dialogues that hold task content constant while varying linguistic cues \cite{zhou_trusting_2019, volkel_user_2022, volkel_examining_2021}. For instance, \citet{volkel_user_2022} created extraverted, introverted, and neutral chatbot variants by hand-authoring sets of fixed dialogue scripts in which the informational content was identical, but the phrasing was systematically varied (e.g., concise vs. verbose responses, use of emojis and exclamation marks, self-disclosure) to signal different levels of extraversion. Even when natural-language understanding is introduced, the broader interaction remains scripted \cite{ruane_user_2020}. These approaches, while offering precise control over personality expression, are fundamentally limited by their reliance on predefined interaction patterns.

Furthermore, prior research has predominantly focused on single-trait manipulations, particularly the social dimensions of Extraversion and Agreeableness, as these are highly perceptible in interaction \cite{volkel_user_2022, ruane_user_2020, volkel_examining_2021, shumanov_making_2021}. For instance, Extraversion has been manipulated by varying verbosity, emoji use, and language formality to create ``introverted'' versus ``extraverted'' agents \cite{volkel_user_2022, ruane_user_2020, zhou_trusting_2019}. Users reliably distinguish such extremes, generally finding extraverted agents more sociable and sometimes more enjoyable \cite{volkel_user_2022, ruane_user_2020}, although creating a convincingly introverted chatbot via text alone remains difficult \cite{volkel_user_2022}. 

Similarly, manipulating Agreeableness affects user affection and perceptions of warmth \cite{volkel_examining_2021}. \citet{volkel_examining_2021} found that users high in Agreeableness particularly appreciated highly agreeable agents, whereas disagreeable agents tended to be less liked by most users, regardless of their own trait level. These studies underscore that personality design has a significant impact on user perceptions.

Despite the Big Five's rich space of possible personality combinations, most HCI studies have examined only one or two traits at a time or focused exclusively on extreme ends of trait spectra \cite{ahmad_extrabot_2020, volkel_user_2022, ruane_user_2020, volkel_examining_2021, zhou_trusting_2019}. This approach leaves a gap in understanding more nuanced or balanced agent personalities. Moreover, determining which personality profile would be optimal for users presents a fundamental design challenge.

\subsection{Personality Alignment Between Users and CAs}
\label{subsec:personality-alignment}

Building on evidence that CA personality systematically influences user perceptions, research has investigated whether personality alignment optimizes user experience. Inspired by social psychology's "similarity–attraction" paradigm, studies demonstrate that users often respond more positively to computer personalities across modalities that resemble their own \cite{moon_how_1996, shamekhi_exploratory_2017, bernier_similarity-attraction_2010, braun_at_2019, lee_designing_2003, nass_does_2001, andrist_look_2015}. Early evidence by \citet{nass_does_2001} showed that matching a user's personality yielded higher liking and perceived performance. \citet{shumanov_making_2021} reported that users were more engaged and willing to make purchases when a customer service agent's extraversion level mirrored their own. Similarly, \citet{gnewuch_effect_2020} found that users disclosed more personal information when an agent's dominance (assertiveness versus submissiveness in communication style) aligned with their own level of dominance. More recently, \citet{amin_kuhail_assessing_2024} tested alignment across three traits in an advisor CA (extraversion, agreeableness, conscientiousness) and found significant effects mainly for extraversion alignment, suggesting this may be the ``primary'' alignment dimension for influencing user perceptions.

However, some studies challenge the universality of this similar\-ity-attraction effect in HCI contexts \cite{isbister_consistency_2000, spagnolli_similarity_2025}. \citet{isbister_consistency_2000} observed that while users recognize and prefer consistent personality cues in an interface agent, they unexpectedly tended to favor a character with a complementary personality over one mirroring their own. Recent work by \citet{spagnolli_similarity_2025} directly tested personality convergence in a text-based health assistant, finding no significant benefit of matching the agent's Big Five profile to the user's. Furthermore, \citet{amin_kuhail_assessing_2024} found no strong impact when separately matching on conscientiousness or agreeableness, indicating that the importance of specific trait alignment may be context-dependent \cite{volkel_examining_2021, gnewuch_effect_2020, spagnolli_similarity_2025}.

These mixed results stem partly from methodological limitations in prior work. Research on personality alignment has been fragmented, typically using categorical matching approaches that examine only one trait at a time \cite{volkel_examining_2021, gnewuch_effect_2020}. The lack of continuous measures of personality alignment and the contradictory findings demand a deeper inquiry into the effects of personality alignment on user perceptions. However, progress has been constrained by the same technical limitations that have limited personality expression research: the difficulty of implementing precise, multi-trait personality control with traditional scripted approaches. 

\subsection{Inducing Personality with Large Language Models}
\label{personality_prompting}

LLMs fundamentally transform CA personality research by enabling systematic personality expression through natural language prompting rather than rigid, pre-programmed responses. This technological shift addresses the core constraints that have limited both personality expression and alignment research. \citet{kovacevic_chatbots_2024} note that while fine-tuning can be effective, it is resource-intensive, whereas prompt engineering scales more readily in applied CA design. Most LLM applications, including CAs studied in HCI, are built on closed-source models accessed via APIs \cite{pang_understanding_2025}. A recent CHI meta-review shows that 130 of 153 CHI 2020–2024 papers that engaged with LLMs (84.98\%) used closed-source GPT-family models \cite{pang_understanding_2025}. In such settings, access to the underlying model parameters for fine-tuning is infeasible, making personality prompting techniques the primary approach for injecting personality into LLM-based CAs.

Within this prompting paradigm, two primary techniques have emerged for steering personality. The first is persona-based prompting, where LLMs are prompted to role-play a person or social identity via short profiles or richer narrative backstories, enabling outputs with a targeted personality or viewpoint \cite{moon_virtual_2024, liu_evaluating_2024, deshpande_toxicity_2023}. Subsequent research has examined the ethical and methodological properties of this persona-based prompting. \citet{haxvig_ive_2025} reported that, using persona-based prompting, LLMs drifted toward gendered occupational stereotypes, repeatedly associating women with creative roles such as graphic designers or illustrators. Similarly, \citet{liu_evaluating_2024} found that LLMs can revert to stereotypical stances when prompted with incongruous persona attributes, and \citet{deshpande_toxicity_2023} and \citet{wan_are_2023} observed that persona assignment can substantially increase harmfulness and toxic agreement in conversations. Related concerns around bias, inclusivity, and the difficulty of governing dynamically generated personas have also been raised by the HCI community \cite{desai_personas_2025, dubiel_voicecraft_2024, zargham_designing_2024}. Additionally, \citet{hu_quantifying_2024} found that persona variables often explain limited variance across many subjective tasks. Given these biases and the limited control afforded by persona prompting, recent research has increasingly positioned adjective-based personality prompting as a more controllable alternative for shaping CA behavior, grounded in psychometric literature, by specifying directly with trait adjectives such as ``talkative'' and ``cooperative'' \cite{goldberg_development_1992}.

The adjective-based personality prompting approach has already begun reshaping CA implementations, with earlier work demonstrating the viability of this approach \cite{alessa_towards_2023, chen_llm-empowered_2023, irfan_between_2023,kovacevic_chatbots_2024, shahid_exploring_2025}. \citet{ramirez_controlling_2023} induced Big Five-styled outputs through discrete prompts using trait adjective and generate-and-rank procedures, demonstrating controllable style while preserving semantic fidelity, operating at polarized endpoints of traits. \citet{jiang_evaluating_2023} similarly showed that LLMs can be prompted toward high or low poles of specific traits using adjective-based prompting, validating the general feasibility of adjective-based personality control. These studies established single-trait, binary-level control as a foundation for personality prompting research.

More recent advances have addressed the multi-trait configuration. \citet{jiang_personallm_2024} moved from single traits to a simultaneous multi-trait configuration using similar adjective-based prompting. PersonaLLM's BFI-44 scores were consistent with assigned profiles, with the framework discretizing each trait to two endpoints \cite{jiang_personallm_2024}. \citet{sutcliffe_survey_2023}'s survey further characterizes this landscape, showing that many personality specifications commonly used in CAs are inherently discrete, which structurally encourages endpoint configurations.

Research on user preferences has also examined the desirability of different personality intensities. \citet{volkel_examining_2021} conclude, ``our findings point to a need for moderate instead of extreme chatbot personalities.'' Addressing medium-level personality expression in LLM-based CAs, \citet{serapio-garcia_personality_2025} introduce a more psychometrically grounded prompting framework that shapes each Big Five trait at nine graded levels using 104 trait adjectives. Their results demonstrate reliable single-trait nine-level control and concurrent shaping under extreme settings, with concurrent evaluation focusing on the lowest and highest intensities.

Investigating whether CA personality should match user personality requires configuring all five traits simultaneously. Prior studies have been limited to examining one trait at a time \cite{volkel_examining_2021, gnewuch_effect_2020}, but if alignment effects depend on holistic profile similarity rather than isolated trait matches, as the mixed findings reviewed in Subsection \ref{subsec:personality-alignment} suggest, then studying these effects demands concurrent medium-level control across the full Big Five. While \citet{serapio-garcia_personality_2025} demonstrate concurrent shaping, their evaluation focuses only on the lowest and highest intensities, leaving open whether stable medium-level expressions can be achieved when all five traits are shaped simultaneously.

This reveals two interrelated gaps. First, no existing prompting framework has demonstrated concurrent medium-level control across all Big Five traits, which is a prerequisite for studying holistic personality alignment. Second, we lack evidence of how LLM-based CAs’ personality affects user perceptions in applied contexts where CAs are designed to help users accomplish specific goals.

\subsection{Performing Goal-Oriented Tasks with CAs}
Goal-oriented CAs are designed to help users complete concrete tasks through dialogue, with interactions structured around clear subtasks and measurable outcomes \cite{peng_soloist_2021, zhang_dialogpt_2020}. Recent analyses of large-scale ChatGPT usage show that task-focused exchanges are central to real-world use. In the OpenAI study, two topic categories associated with goal-directed interaction, Practical Guidance and Information Seeking, together accounted for 53.2\% of all user messages \cite{chatterji_how_2025}. These categories include activities such as planning travel or study schedules, searching for information, troubleshooting technical problems, producing summaries or drafts, and receiving step-by-step guidance across everyday domains. Although such dialogs emphasize efficiency, prior work shows that the human-like qualities of the agent can meaningfully influence perceptions of Enjoyment and Trust, which in turn shape satisfaction even in transactional contexts \cite{haresamudram_tasks_2025}.

Researchers often choose travel scenarios for CA studies because they provide concrete goals and decisions (e.g., book a hotel, find activities) without the high-stakes concerns of domains like banking or healthcare \cite{haresamudram_tasks_2025}. This domain enables controlled experimentation through standardized vignettes, ensuring that every participant faces the same context and subtasks. \citet{atzmuller_experimental_2010} describe how vignette-based experiments frame consistent situations for all users, reducing variability from personal context differences. Recent CA studies have employed travel assistant vignettes to control for context and focus participants on comparable subtasks (e.g., lodging, attractions) in short trip-planning exercises \cite{haresamudram_tasks_2025, nguyen_user_2022}.

Within this domain, researchers have identified several factors that influence user perceptions and task outcomes. \citet{haresamudram_tasks_2025} demonstrated that interaction modality and communication medium independently influence user perceptions in travel-assistant contexts, with text-based interactions yielding higher Trust and Anthropomorphism ratings than voice-only alternatives. \citet{nguyen_user_2022} found that users approach tasks differently with free-form chatbot interfaces versus structured menu systems, which affects both efficiency and satisfaction. Additionally, \citet{yeh_how_2022} noted that guidance strategies must be carefully timed and designed—giving an example query at the task's start or offering hints after errors can reduce user confusion and improve completion rates, though excessive upfront rules may frustrate users.

Taken together, this stream of work in goal-oriented settings has largely emphasized interaction medium and modality as the primary levers shaping user perceptions and outcomes \cite{nguyen_user_2022, haresamudram_tasks_2025}. However, we lack evidence on how agent personality affects user perceptions when the task and interface are held constant. This represents a significant gap, as personality may offer an additional design dimension for optimizing user experience in goal-oriented conversational systems. The convergence of systematic personality control capabilities with realistic interaction contexts creates an opportunity to bridge decades of theoretical insights about personality effects with practical implementation in applied settings.

\section{Study}
We employed a $3\times 1$ between-subjects design (N=150, $n=50$ per condition) to examine how CA personality expression level, operationalized at three levels (low, medium, high), shapes user perceptions during a goal-oriented task. Participants were assigned to one of the three conditions in a counter-balanced order. Specifically, those in the low condition interacted with Low CA, those in the medium condition with Medium CA, and those in the high condition with High CA—each configured to exhibit the corresponding level of personality expression across all Big Five traits. Holding the task constant, the New York City trip-planning scenario enabled direct cross-condition comparisons while minimizing task-induced variance.

The CA was built using GPT-4o, OpenAI’s non-reasoning flagship model at the time of the study (gpt-4o-2024-08-06), selected for its strong performance in instruction-following during our internal validation (see details in Appendix \ref{appendix:eval}), and based on evidence from prior studies demonstrating the effectiveness of OpenAI’s LLMs for personality prompting \cite{serapio-garcia_personality_2025, jiang_personallm_2024}.

\subsection{Task Design}
We adapted our task from prior work by \citet{haresamudram_tasks_2025}, who employed a vignette-based, goal-oriented travel assistant scenario. This task was chosen because vignettes fix the user role, goal, and constraints before interaction, thereby reducing nuisance variance from idiosyncratic task interpretations \cite{atzmuller_experimental_2010}.

\subsubsection{NYC Trip Planning Task}
\label{sec:nyc-trip}
Participants were told they would plan a one-day summer trip to New York City with our CA. The interaction was capped at 10 minutes and adapted from prior travel-assistant vignettes \cite{haresamudram_tasks_2025, nguyen_user_2022}, but implemented entirely via open-ended chat with following subtasks: (1) discover your top priority (what matters most when you travel, such as budget, food, or sights), (2) choose a base neighborhood in NYC, (3) select a rest stop or day-use accommodation, (4) plan your cultural or recreational activities, (5) decide on within-city transportation, (6) review the day’s plan, and (7) generate a one-day summer trip itinerary with a complete, time-stamped schedule from morning to night. These subtasks provided the discrete objectives and measurable outcomes characteristic of goal-oriented interactions.

\subsubsection{Task Interface}
\label{sec:task-interface}
\begin{figure*}[ht]
\centering
\includegraphics[width=\textwidth]{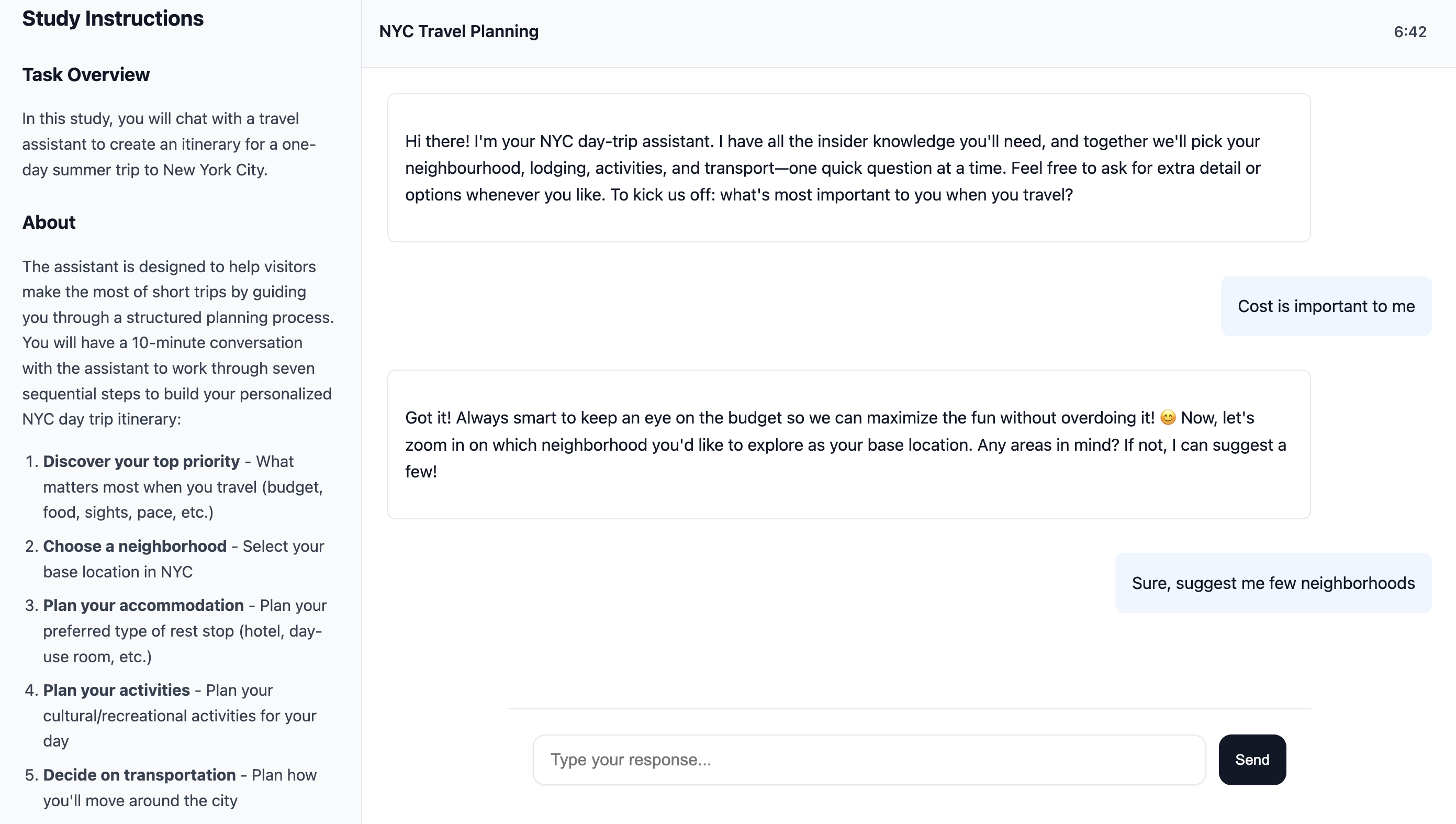}
\caption{Text-only study interface and example exchange with the NYC day-trip CA. The left sidebar presents the vignette and the planning steps accessible to participants during interaction with the CA; the right pane displays the chat interface.}
\Description{Screenshot of a two-column web interface. The left column, titled “Study Instructions,” presents a Task Overview and numbered steps for planning a one-day NYC trip (identify priorities, choose a neighborhood, plan accommodation and activities, and decide transportation). The right column, titled “NYC Travel Planning,” shows a chat transcript: the CA greets the user, explains it will guide planning one step at a time, and asks what matters most when they travel; the user replies, “Cost is important to me.” The CA acknowledges in a friendly tone with an emoji, then asks which neighborhood to use as a base and offers to suggest options. The user responds, “Sure, suggest me few neighborhoods.” A text input field and “Send” button appear at the bottom.}
\label{fig:interface}
\end{figure*}

We used a text-only chat interface (see Figure \ref{fig:interface}) to maximize experimental control and minimize confounds. Participants communicated through free-text input, and the CA responded in natural language; we deliberately excluded buttons, quick replies, menus, avatars, and voice interactions. By standardizing both modality (free text) and medium (text) \cite{haresamudram_tasks_2025}, we isolated personality manipulation as the primary source of variance in user perceptions.

To balance user autonomy with task structure in the absence of UI scaffolding, the CA employed lightweight guidance strategies \cite{paranjape_human-like_2021}. These included affirming user contributions and offering brief exemplars, a technique previously validated in travel-assistant settings \cite{haresamudram_tasks_2025}. This implementation enabled the CA to express personality through textual replies while maintaining consistent task progression across all experimental conditions.

\subsection{Conversational Agent Design and Prompting Framework}
\subsubsection{Personality Prompting: Trait Modulation Keys (TMK)}
To address the gaps identified in Subsection \ref{personality_prompting} and the need for a controllable personality for our NYC trip-planning CA, we develop TMK: a modular personality-prompting framework that enables simultaneous control across all Big Five traits at multiple expression levels, including medium-level expression, primarily using trait adjectives without relying on personas.

TMK operates through independent (\textit{trait, level}) controls applied concurrently across Big Five traits. Each trait can be tuned to Low, Medium, or High expression levels rather than only extremes, a capability that, to our knowledge, represents the first systematic approach to shape medium-level expressions in all Big Five personality traits concurrently for LLM-based CAs, extending the previous personality prompting frameworks \cite{jiang_personallm_2024, serapio-garcia_personality_2025, jiang_evaluating_2023}. For example, TMK can construct an LLM personality that is low in Openness and Emotional Stability, high in Conscientiousness and Agreeableness, and medium in Extraversion, all within a single prompt. This capacity to define and combine trait intensities enables CAs with fine-tuned stylistic alignment and behavior patterns across roles.

The framework implements each personality configuration thro\-ugh two complementary keys for every (\textit{trait, level}) Cartesian pair:

\begin{itemize}
\item \textbf{Personality Key.} A concise, psychometrically grounded description of the agent's intended trait intensity. For Low and High, these are derived from validated psychometric resources \cite{saucier_mini-markers_1994, goldberg_development_1992, mccrae_introduction_1992}. The high and low-level descriptors mostly use unipolar adjectives empirically linked to each trait. Unlike bipolar adjectives (``friendly–unfriendly''), which can introduce ambiguity, unipolar terms (``kind'', ``creative'') allow precise mapping to trait presence. Due to a lack of direct lexical markers for the medium level in prior research, we derived mid-range descriptors from previous research on behavioral descriptions of individuals who exhibit flexible or average trait levels \cite{grant_too_2011, grant_rethinking_2013, carter_extreme_2018}. All descriptors employ the ``You...'' prefix to leverage improved role adherence documented in instruction-following LLMs \cite{ouyang_training_2022, zhou_context-faithful_2023, white_prompt_2023}.

\item \textbf{Style Cues Key.} A set of targeted linguistic directives that govern the agent's tone, lexical choices, syntactic constructions, and discourse markers to align surface-level communication with the intended trait. Each style cue is informed by prior psycholinguistic research on how personality traits manifest in language \cite{pennebaker_linguistic_2001,yarkoni_personality_2010, mairesse_using_2007,schwartz_personality_2013, campbell_bodily_1978, carment_persuasiveness_1965, mehl_personality_2006, patterson_social_1966, rutter_visual_1972, doi:10.1177/0261927X12460844}. These style cues enable trait expressions to persist outside of psychometric questionnaire formats, addressing concerns about uneven personality visibility in previous personality prompting frameworks in downstream tasks (e.g., writing a blog post) \cite{jiang_personallm_2024, serapio-garcia_personality_2025}.
\end{itemize}

This dual-key architecture separates behavioral stance (e.g., risk-taking, diligence, social warmth) from linguistic surface form (tone, hedging, pace, punctuation), ensuring trait-consistent behavior while maintaining communicative flexibility. All 30 Personality Keys and Style Cues Keys are available as supplementary material, including 15 Personality Keys and 15 Style Cues Keys, with three keys per trait representing three levels of expression.

\begin{figure*}[ht]
\centering
\includegraphics[width=\textwidth]{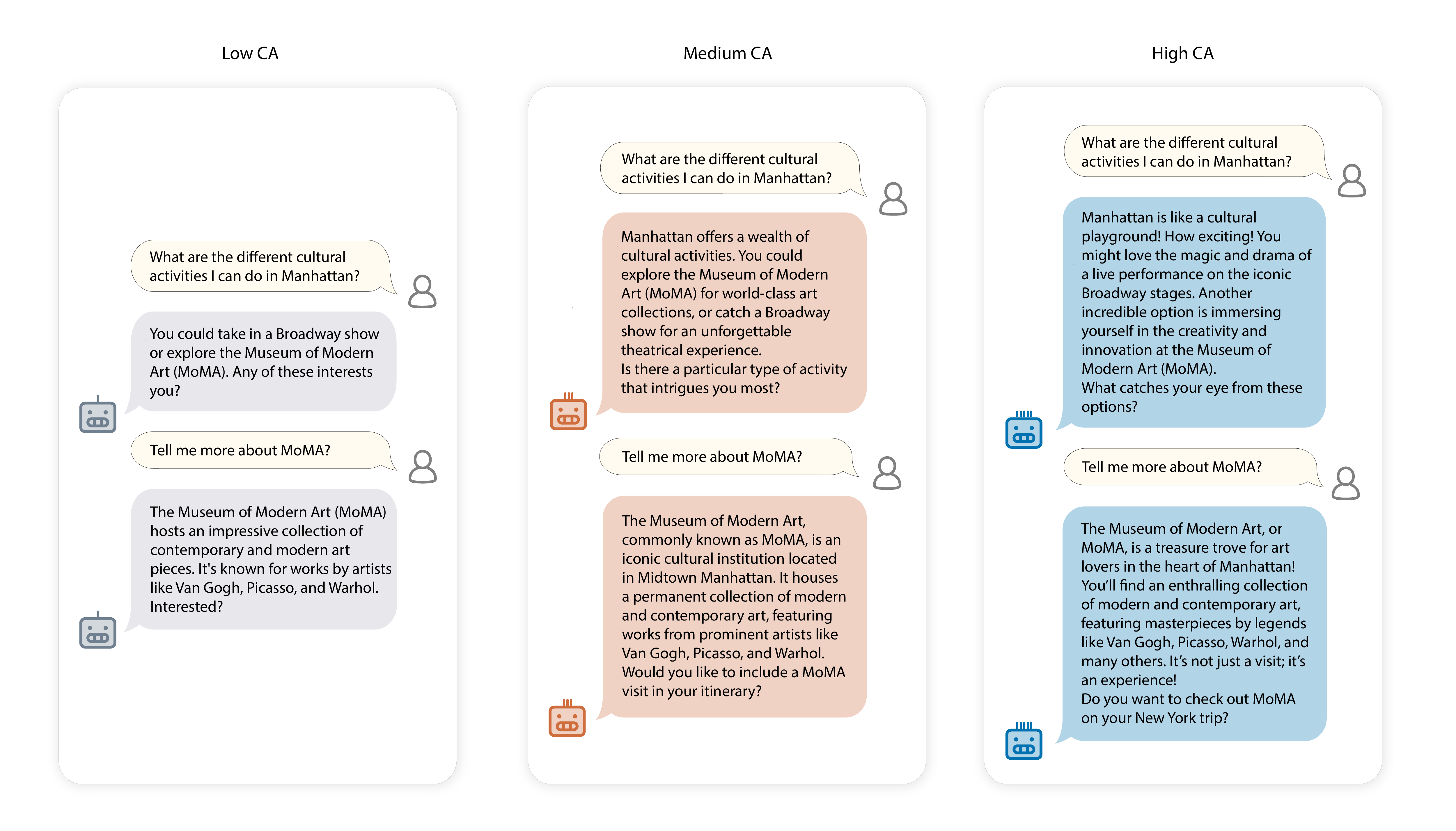}
\caption{TMK prompting framework used to elicit three distinct personality profiles, illustrated with example responses from Low CA, Medium CA, and High CA.}
\Description{Three side-by-side chat mockups compare CA styles. Left panel, labeled “Low condition CA,” shows gray chat bubbles with a robot icon; the CA answers briefly, suggests Broadway or the Museum of Modern Art (MoMA), and provides a short, factual MoMA description without emotion. Middle panel, labeled “Medium condition CA,” uses orange bubbles; the CA offers the same options with a warmer, fuller paragraph and a gentle follow-up question (“Is there a particular type of activity…?”), then provides a clear, concise MoMA summary and a polite offer to include it in the itinerary. Right panel, labeled “High condition CA,” uses blue bubbles with exclamation marks and vivid language (“cultural playground,” “treasure trove,” “It’s an experience!”), ending with an eager call to action about visiting MoMA. User speech bubbles are identical across panels, emphasizing that only the CA’s personality expression—not the task content—changes.}
\label{fig:conditions}
\end{figure*}

Following our $3 \times 1$ between-subjects design, we instantiated three uniform CA profiles: (i) \textbf{Low CA}: all five traits set to low expression; (ii) \textbf{Medium CA}: all five traits set to medium expression; and (iii) \textbf{High CA}: all five traits set to high expression. 
This design choice reflects both theoretical and psychometric considerations, as traditional Big Five profiles in the population typically show coherent patterns across traits, and pronounced divergences are atypical and occur only in a minority of individuals \citep{mccrae_cross-cultural_2002, goldberg_development_1992, park_meta-analytic_2020, allik_unusual_2018}. Empirically, Big Five traits do not operate as independent sliders in human psychology; rather, they tend to show small to moderate positive intercorrelations rather than strict orthogonality when Neuroticism is interpreted as Emotional Stability \citep{wilt_big_2019, musek_general_2007, van_der_linden_general_2010}. These intercorrelations give rise to higher-order metatraits such as Stability, which captures the shared variance among Conscientiousness, Agreeableness, and Emotional Stability; Plasticity, which reflects covariance between Extraversion and Openness; and the general factor of personality (GFP), which represents the common variance shared across all five traits \citep{musek_general_2007, van_der_linden_general_2010}. By employing three uniform CA profiles that move all five traits together, we thus ensure ecologically valid and stable personality expression profiles. See Figure \ref{fig:conditions} for illustrative CA responses produced via TMK across Low, Medium, and High profiles.

\subsubsection{Personality Prompt Validation}
\label{sec:validation}

Before deploying TMK in our user study, we recognized that effective personality manipulation requires precise control and generalizability beyond specific configurations. Therefore, we conducted comprehensive validation across all possible 243 ($3^5$, three levels and five traits) combination spaces to ensure TMK's reliability and steerability in accurately shaping intended personality expressions across all possible configurations. This comprehensive validation served dual purposes: to confirm that our three experimental conditions would produce distinct, measurable personality profiles, and to establish TMK's broader applicability for future personality-aware CA design. We instantiated all 243 TMK combinations and queried GPT-4o with two standard Big Five inventories, the 20-item Mini-IPIP \cite{donnellan_mini-ipip_2006} and the 44-item BFI-44 \cite{john_big-five_1999}, plus four external criteria sets: Positive and Negative Affect Schedule (PANAS) \cite{watson_development_1988}, Buss and Perry Aggression Questionnaire (BPAQ) \cite{kiewitz_aggression_2007}, Portrait Value Questionnaire-Revised (PVQ-RR) \cite{schwartz_refining_2012}, and Short Scale of Creative Self (SSCS) \cite{karwowski_big_2013}. Similar inventory-based validation strategies have been used in prior works, for example, by \citet{jiang_personallm_2024}, \citet{serapio-garcia_personality_2025}, and \citet{salecha_large_2024}. In addition, \citet{huang_reliability_2024} showed that different LLMs produce consistent responses on a personality inventory, offering relevant context for our methodological choices. Responses on 1--5 Likert scales yielded average trait scores, which we mapped into three equal fixed bands, Low [1.00, 2.33), Medium [2.33, 3.67], High (3.67, 5.00], to assess control fidelity. Building on validation approaches from prior works \cite{serapio-garcia_personality_2025, jiang_personallm_2024}, our assessment encompassed three dimensions.

\paragraph{Convergent \& Discriminant Validity} We evaluated convergent and discriminant validity using a Multi-Trait Multi-Method approach, comparing trait scores from the Mini-IPIP and BFI-44. Convergent validity was assessed by correlating the same trait across instruments, while discriminant validity required that each Mini-IPIP trait correlate weakly with the four non-matching BFI-44 traits. We quantified separation as $\Delta = \text{(convergent } r) - \text{(mean non-target } r)$. TMK-shaped responses yielded correlations between matched traits across instruments (Mini-IPIP and BFI-44) with an average of $\bar{r} = .989$, and correlations with non-matching traits of up to $.28$. The resulting average separation score, $\Delta = .922$, indicates that TMK-targeted trait outputs corresponded with an independent instrument's measure of the same construct and remained distinct from non-target traits.

\paragraph{Reliability} To assess whether traits maintain internal consistency under simultaneous five-trait control, we used three standard coefficients: Cronbach's $\alpha$, McDonald's $\omega$, and Guttman's $\lambda_6$. Under TMK shaping, all five Mini-IPIP traits achieved \emph{excellent} internal consistency for GPT-4o: $\alpha \ge .973$, $\omega \ge .971$, and $\lambda_6 \ge .969$. These results indicate that the personality scales remain highly dependable even under simultaneous five-trait control. 

\paragraph{Criterion Validity} 
Trait–criterion relationships were assessed against theoretical expectations linking Big Five traits to non-personality outcomes, e.g., Extraversion should positively correlate with PANAS Positive Affect and negatively with Negative Affect \cite{burger_personality_2000}; Agreeableness should correlate negatively with BPAQ aggression facets \cite{jiang_how_2022}; Conscientiousness should correlate positively with PVQ-RR Achievement, Conformity/Tradition, and Security \cite{parks-leduc_personality_2015, vecchione_five_2023}; Emotional Stability should correlate negatively with Anger/Hostility \cite{quan_relationship_2024}; and Openness should correlate positively with creative self-beliefs \cite{pavlic_relationship_2023}. Under TMK shaping in GPT-4o, Extraversion correlated positively with PANAS Positive Affect ($r = .87$) and showed a ($r = -.01$) relationship with Negative Affect. Agreeableness was negatively related to BPAQ aggression facets—Physical Aggression ($r = -.83$), Verbal Aggression ($r = -0.87$), Anger ($r = -.03$), and Hostility ($r = -.39$). Conscientiousness tracked value endorsements on the PVQ-RR: Achievement ($r = .12$), Conformity/Tradition ($r = .54$), and Security ($r = .95$). Emotional Stability was negatively associated with BPAQ Anger ($r = -.98$) and Hostility ($r = -.89$). Finally, Openness aligned with creative self-beliefs measured by the SSCS: Creative Self-Efficacy ($r = .96$) and Creative Personal Identity ($r = .97$). These criterion relationships demonstrate that TMK-shaped personalities exhibit theoretically consistent external correlates beyond mere self-report consistency.

\begin{figure*}[ht]
\centering
\includegraphics[width=\textwidth]{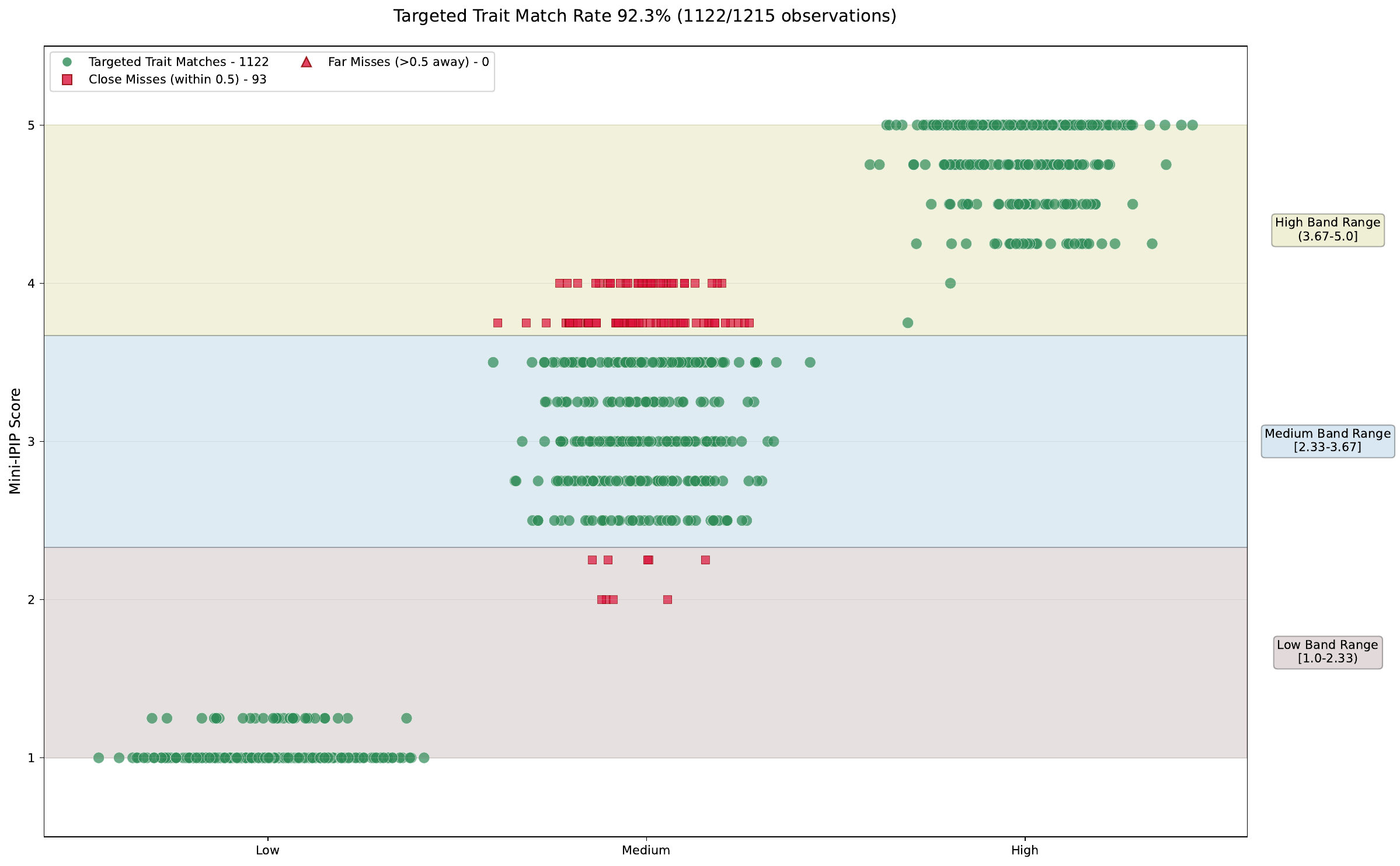}
\caption{Control fidelity of TMK personality prompting on the CA, measured with Mini‐IPIP across all 243 trait–level configurations. Each point is a trait–level observation ($n=1215$). Shaded bands mark target ranges: Low = $[1.0,2.33)$, Medium = $[2.33,3.67]$, High = $(3.67,5.0]$. Green circles denote targeted trait matches ($1122$; 92.3\%), red squares are close misses (within 0.5; $93$), and red triangles are far misses ($0$). Misses occur only for Medium targets; Low and High achieve 100\% matches, yielding three clearly separated clusters.}
\Description{Scatter plot titled “Targeted Trait Match Rate 92.3\% (1122/1215 observations).” The x-axis shows three categories (Low, Medium, High) and the y-axis shows Mini-IPIP scores from 1 to 5. The background is partitioned into three horizontal shaded bands labeled at the right: Low Band Range [1.0–2.33), Medium Band Range [2.33–3.67], and High Band Range (3.67–5.0]. Green circles (matches) cluster near 1.0–1.3 for Low, 2.7–3.4 for Medium, and 4.2–5.0 for High. Red squares (close misses) appear only in the Medium category near 2.0–2.2 and 3.7–4.0. No red triangles (far misses) are present. A legend at the top-left explains the marker types and counts.}
\label{fig:targeted_trait}
\end{figure*}

\paragraph{Control Fidelity} Two forms of control fidelity were assessed. First, trait–level correlation was evaluated by correlating the targeted level for each trait (Low = 1, Medium = 2, High = 3) with the resulting Mini-IPIP trait scores using Spearman's $\rho_s$. The average correlation across traits was $\bar\rho_s = .97$ ($p < .001$), indicating significant alignment between targeted levels and resulting scores. Second, a Targeted Trait Match assessment was conducted across all 1215 trait–level observations (243 combinations × 5 traits), classifying responses as matches when the resulting Mini-IPIP trait scores for each targeted trait level fell within the predefined target band and misses otherwise. This analysis revealed a 92.3\% targeted trait match rate (1122/1215 observations). Notably, using TMK, the medium level proved most challenging to shape precisely, with all 93 misses originating exclusively from medium-targeted traits whose resulting Mini-IPIP trait scores fell just outside their intended band boundaries. However, these deviations represented close misses; none exceeded half a Likert point from the target range, and no medium-targeted trait scores migrated to the extremes of low or high expression levels. Despite these challenges, the resulting Mini-IPIP trait score distributions formed visually distinct clusters across low, medium, and high levels (see Figure \ref{fig:targeted_trait}). 

To further validate the fidelity of our three CA profiles, we ran the Mini-IPIP 40 times for each experimental condition, yielding 200 trait–level observations per condition. The number of runs was determined a priori: at least 39 runs were required to achieve 80\% power ($\alpha=.05$, $1-\beta=.80$). In the medium condition, 180 of 200 observations (90.0\%) fell within their target bands. The low and high conditions yielded 200 of 200 matches (100\%). These three condition-level results confirm that TMK produced stable and distinct CA profiles, ensuring consistent behavioral expression across the interaction sequence.

This comprehensive validation confirmed TMK's capacity to reliably shape distinct, measurable personality profiles in LLMs aligned with specified target profiles. 

\subsubsection{Prompt Architecture}
The CA's prompt architecture is modular and condition-aware, assembling personality-driven behavioral directives alongside structured task parameters in a controlled sequence. The prompt architecture first establishes the CA's role as a New York City trip-planning assistant. It binds behavioral decisions to predefined ``Personality Key'' and communication patterns to ``Style Cues Key'', both determined by the assigned experimental condition. It then integrates task instructions across multiple subsections: (i) Goal (planning objectives and success criteria); (ii) Prompt Construction Rules (comprehensive task specifications plus structured questionnaires for information elicitation); (iii) Conversation Flow (exemplar exchanges and turn-taking protocols); (iv) Chat Termination Guidelines (completion criteria tied to trip itinerary generation); and (v) Restrictions (behavioral and content constraints to preserve experimental validity). Appendix \ref{appendix:prompts} details the complete prompt structure and one example prompt. This design cleanly separates personality manipulations from procedural scaffolding while ensuring personality expressions remain active throughout the dialogue, thereby enabling precise experimental control and consistent behavioral expression across the interaction sequence.

\subsection{Measures}
\subsubsection{Personality Measure (Mini-IPIP)}
\label{sec:measures}
We measured participants' personality profile using the 20-item Mini-IPIP \cite{donnellan_mini-ipip_2006}, which includes four items per Big Five domain rated on a 5-point Likert scale (1 = strongly disagree, 5 = strongly agree). This brief measure demonstrates strong psychometric properties, with internal consistencies of $\alpha = .65$–$.77$ and strong convergence with the 50-item parent scales ($r = .85$–$.93$) \cite{donnellan_mini-ipip_2006}.

For scoring, negatively keyed items were reverse-scored, and each domain score was calculated as the mean of its four items. Consistent with the view that Emotional Stability is the opposite pole of Neuroticism, we report Emotional Stability as  ($6 - Neuroticism$) \cite{donnellan_mini-ipip_2006}. Following standard Big Five labeling, we refer to the Mini-IPIP ``Intellect/Imagination'' domain as ``Openness to Experience'' or ``Openness'' in short \cite{john_big-five_1999, donnellan_mini-ipip_2006}.

\subsubsection{User Perception Measures}
To evaluate user perception of the CA, we use six measures covering Intelligence, Enjoyment, Anthropomorphism, Intention to Adopt, Trust, and Likeability based on \citet{10.1145/3571884.3597139}'s framework linking user perceptions to agent relationships.

The Intelligence, Anthropomorphism, and Likeability subscales from the Godspeed Measures \cite{bartneck_godspeed_2023, bartneck_measurement_2009} were employed using 5-point semantic differentials. The Enjoyment was measured with three items adapted from \citet{moussawi_how_2021} on a 7-point Likert scale, evaluating whether users found the interaction with the CA enjoyable, interesting, and fun. Intention to Adopt was measured with two items adapted from \citet{moussawi_how_2021} on a 7-point Likert scale, assessing users' plans to start using and experiment with the CA within the coming months. Trust was also measured on a 7-point Likert scale adapted from \citet{jian_foundations_2000}. The questionnaires can be found in Appendix \ref{appendix:questionnaires}.

In addition to these measures, we included three attention check questions. These were simple instruction-following items and were interspersed with the main measures in randomized order to minimize response bias.

\label{sec:user-perception}
\subsubsection{Personality Alignment Measures}
We quantify alignment between each participant's Big Five profile and the CA using (i) a multivariate ``Personality Alignment Score'' and (ii) ``Individual Trait Distance'' for each trait. 

Let $\mathbf{P}=(P_O,\allowbreak P_C,\allowbreak P_E,\allowbreak P_A,\allowbreak P_{ES}) \in [1,5]^5$ denote the participant's trait scores (Mini-IPIP subscale means, and let $\mathbf{C}=(C_O,\allowbreak C_C,\allowbreak C_E,\allowbreak C_A,\allowbreak C_{ES})$ denote the CA's trait settings, where $C_i \in \{1,\allowbreak 3,\allowbreak 5\}$ encodes low, medium, and high, respectively. $C_i$ scores are based on the modal observed trait scores in our TMK validation, which were 1 for low, 3 for medium, and 5 for high. These values best represent the most typical realized expression level for each setting (see Section \ref{sec:validation}).

\paragraph{Personality Alignment Score.} We define a normalized, Euclidean-distance-based score:

\begin{equation}
\begin{aligned}
\text{Personality Alignment Score}
&= 1 - \frac{\lVert \mathbf{P}-\mathbf{C} \rVert_2}{\sqrt{80}} \\
&= 1 - \frac{\sqrt{\sum_{i=1}^{5} (P_i - C_i)^2}}{\sqrt{80}}
\end{aligned}
\end{equation}

The denominator $\sqrt{80}$ is the maximum possible Euclidean distance given the scales (for each trait, the largest difference is $|5-1|=4$, so $\sum 4^2=80$). The score is thus bounded in $[0,1]$, where higher values indicate better personality alignment between the participant and the CA. Even though the $C_i$ scores representing each expression level are categorical, the Personality Alignment Score is continuous.

\paragraph{Individual Trait Distance.}
For each trait \(i\) in {Openness, Conscientiousness, Extraversion, Agreeableness, Emotional Stability}, we compute
\[
\text{Trait Distance}_i \;=\; \lvert P_i - C_i \rvert,
\]
the absolute difference between the participant's trait score and the CA's setting on that trait. $\text{Trait Distance}_i$ ranges from $0$ (minimum distance) to $4$ (maximum distance). For example, Openness distance would be \( \lvert P_O - C_O \rvert \), where \(P_O\) is the participant's Openness score and \(C_O\) is the CA's Openness setting.

\subsection{Procedure}
\begin{figure*}[ht]
\centering
\includegraphics[width=\textwidth]{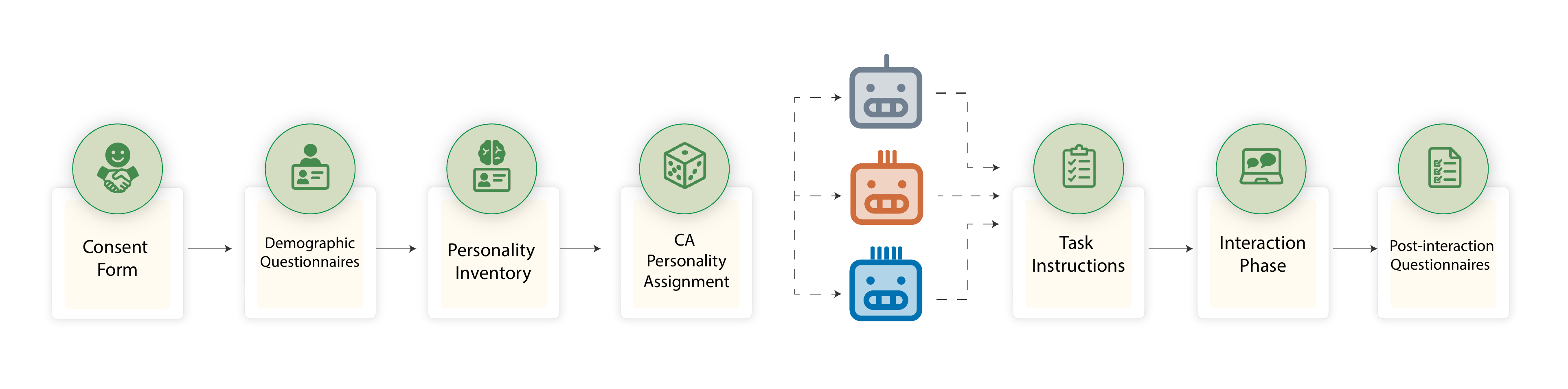}
\caption{Single session study flow}
\Description{Left-to-right flow diagram with seven labeled boxes connected by arrows: “Consent Form” $\rightarrow$ “Demographic Questionnaires” $\rightarrow$ “Personality Inventory” $\rightarrow$ “CA Personality Assignment.” A dashed bracket then encloses three stacked robot icons—gray, orange, and blue—representing the Low, Medium, and High CA conditions. Arrows from each robot converge to “Task Instructions,” then proceed to “Interaction Phase,” and finally to “Post-interaction Questionnaires.”}
\label{fig:procedure}
\end{figure*}

Participants accessed the study through Prolific\footnote{\url{https://www.prolific.com/}} and completed all components within a single session. The procedure is illustrated in Figure \ref{fig:procedure}. Upon entry via the platform link, participants first reviewed and signed an online IRB-approved consent form before providing demographic information, including their prior experience with CAs. Next, they completed the Mini-IPIP personality inventory to capture Big Five trait scores for subsequent alignment analyses (see Section \ref{sec:measures}); these scores were not used for condition assignment. Following the personality assessment, participants were assigned to one of three CA personality profiles (Low CA, Medium CA, or High CA) in counter-balanced order. Before beginning the interaction, participants read detailed task instructions for NYC day-trip planning (see Section \ref{sec:nyc-trip}) that specified their user role, goals, constraints, and the 10-minute time limit, confirming their understanding before proceeding. They were then routed to our custom chat interface (see \ref{sec:task-interface}) where they interacted with the CA to complete the seven planning subtasks. The interaction phase concluded either upon successful completion of all subtasks or expiration of the 10-minute timer, whichever occurred first. Finally, participants completed post-interaction questionnaires (with questionnaire items presented in a randomized order) measuring their perceptions of the CA (see Section \ref{sec:user-perception}). Further, they were given the option to provide qualitative feedback about their interactions with the CA, should they choose to do so. Participants took a median of 14 min 25 s to complete the entire study. After completing these steps, participants were redirected to Prolific for study completion.

\subsection{Participants}
To detect a one-standard-deviation effect in our 3×1 between-subj\-ects design, a priori power analysis indicated a minimum sample of 148 participants (\(\alpha = .05; 1 - \beta = .80\)). To meet this threshold, we therefore recruited 150 US adults through Prolific, with 50 participants assigned to each experimental condition (low, medium, high) in a counter-balanced order.

Recruitment filters required a minimum task approval rate of 98\% and at least one year of activity on the platform. Our eligibility was defined as (i) self-reported high English proficiency for effective CA interaction, and (ii) at least minimal prior exposure to CAs. All participants met these criteria, and all identified English as their native language. Consistent with the inclusion criterion, 72\% described themselves as ``very'' or ``completely'' familiar with CAs (ratings 4--5 on the 5-point familiarity scale), and 71.3\% reported using a CA at least weekly (ratings 4--5 on the frequency scale).

The final sample spanned 18--73 years (\(M = 41.95, SD = 10.91\)). Gender identification was 52.0\% female (\(n = 78\)), 46.7\% male (\(n = 70\)), and 1.3\% non-binary/other or undisclosed (\(n = 2\)).
Participants were compensated at an hourly rate of US \$12.50 per hour, exceeding the US federal minimum wage guidance.

\section{Results}
The 150 participants were evenly distributed across three experimental conditions (Low, Medium, and High CA personality; $n=50$ per condition). Of these, 130 (86.7\%) completed all seven subtasks (outlined in Section~\ref{sec:nyc-trip}) within 10-minute time limit: 44 from the low condition, 41 from the medium condition, and 45 from the high condition. The remaining 20 participants completed the majority of subtasks but did not finish all seven, typically failing to generate a complete itinerary due to time constraints. Participants engaged in a comparable number of conversational turns across conditions ($M = 24.62$ turns, $SD = 8.37$ for low; $M = 23.20$ turns, $SD = 4.61$ for medium; $M = 23.26$ turns, $SD = 4.95$ for high). This ensures that the observed differences in user perceptions are attributable to personality variation rather than differential engagement levels.

\begin{figure*}[ht]
\centering
\includegraphics[width=\textwidth]{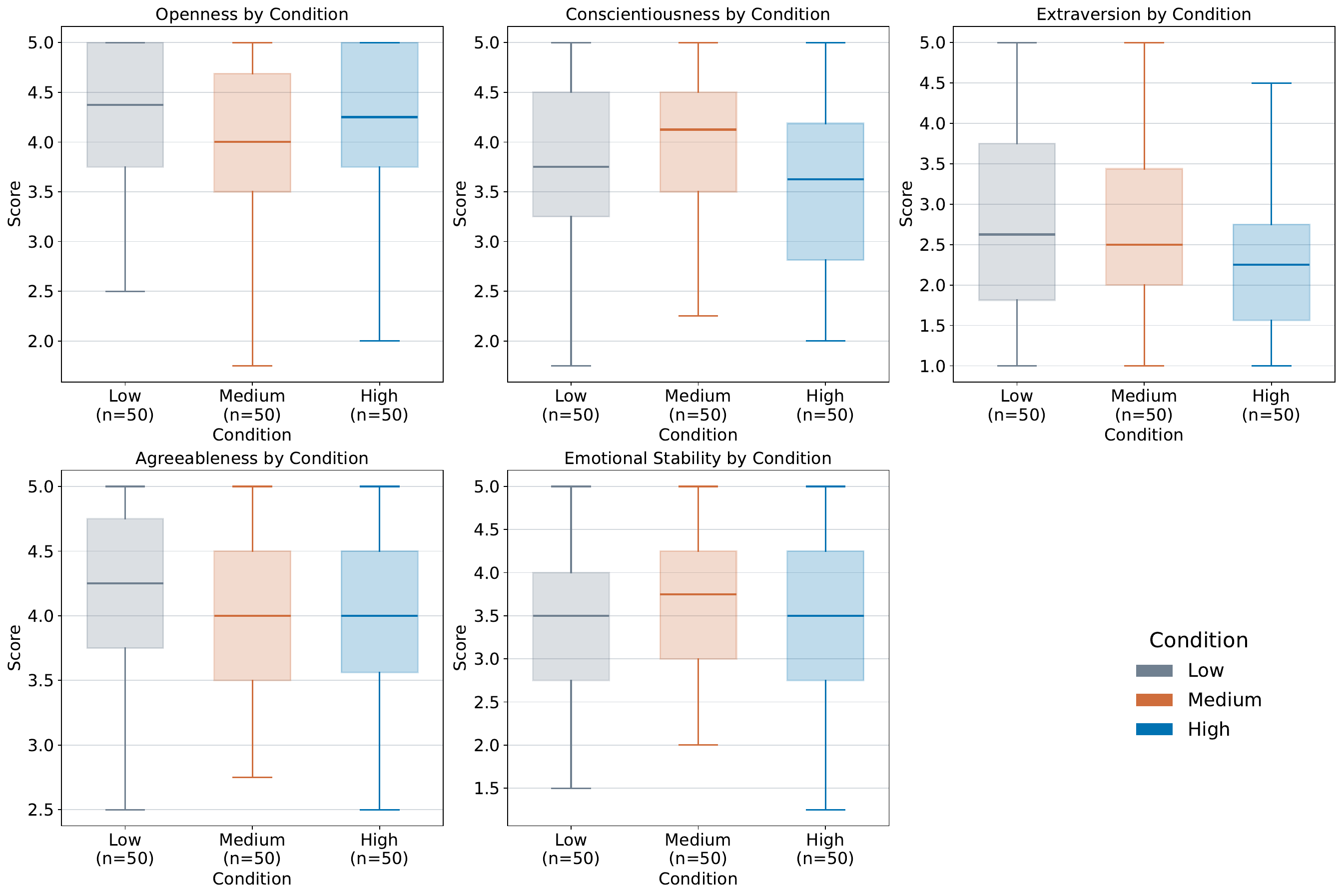}
\caption{Participants' Mini-IPIP Big-Five scores by experimental condition ($n=50$ per condition). Each panel shows boxplots for Openness, Conscientiousness, Extraversion, Agreeableness, and Emotional Stability with three groups (Low, Medium, High). Distributions are broadly comparable across conditions.}
\Description{Five small-multiple boxplot panels, one per Big-Five trait, each with a y-axis from 1 to 5. In each panel, three boxplots—Low (gray), Medium (orange), and High (blue)—summarize participants’ Mini-IPIP scores ($n=50$ per condition). Openness shows similar medians around 4 with overlapping spreads. Conscientiousness shows the Medium condition with a slightly higher median and upper quartile than Low and High. Extraversion medians are low overall, with the Low condition slightly higher than Medium and High. Agreeableness shows medians near 4 with broadly overlapping distributions. Emotional Stability shows the Medium condition with a somewhat higher median than Low, with High in between. A legend at the bottom-right labels the three conditions.}
\label{fig:personality_scores_by_condition}
\end{figure*}

Personality profiles were broadly comparable across experimental conditions (see Figure \ref{fig:personality_scores_by_condition}). Kruskal--Wallis tests indicated no statistically significant differences for Openness ($H(2) = 2.59$, $p = .274$), Extraversion ($H(2) = 3.12$, $p = .210$), Agreeableness ($H(2) = 1.90$, $p = .387$), and Emotional Stability ($H(2) = 2.81$, $p = .246$). Conscientiousness differed across conditions ($H(2) = 6.51$, $p = .039$), with $\mathit{Mdn} = 3.62$ (high), $\mathit{Mdn} = 3.75$ (low), and $\mathit{Mdn} = 4.12$ (medium). Despite this difference, the overall trait distribution remained balanced across groups, reducing the likelihood that individual-difference composition operated as a systematic confound.

At the sample level, descriptive statistics were as follows: Openness ($M = 4.08$, $SD = 0.91$, 95\% CI [3.933, 4.227]), Conscientiousness ($M = 3.73$, $SD = 0.89$, 95\% CI [3.588, 3.876]), Extraversion ($M = 2.63$, $SD = 1.09$, 95\% CI [2.449, 2.801]), Agreeableness ($M = 4.02$, $SD = 0.77$, 95\% CI [3.899, 4.148]), and Emotional Stability ($M = 3.48$, $SD = 0.95$, 95\% CI [3.323, 3.630]).

To further support our use of three uniform CA profiles, we conducted a Principal Component Analysis on participants' Big Five scores (see Table \ref{tab:pca_personality} in Appendix). A single latent dimension (PC1) accounted for 34.3\% of total variance. Extraversion (.57), Agreeableness (.56), and Openness (.51) exhibited strong positive loadings, with Emotional Stability (.30) loading moderately and Conscientiousness (.11) weakly. This pattern is consistent with prior research on higher-order personality structure \citep{musek_general_2007, van_der_linden_general_2010}. These findings empirically corroborate our design choice of uniform CA profiles that shift all five traits in the same direction.

All measures demonstrated good to excellent internal consistency ($\alpha = .81$--$.95$). All variables violated normality assumptions (Shapiro--Wilk tests, $p < .05$); hence, nonparametric analyses were used throughout. All the data is available as supplementary material.

\subsection{Effects of CA Personality (RQ1)}
Results indicate that personality profiles of the CA systematically shape user perceptions across multiple dimensions. Table \ref{tab:means_sds} summarizes means and standard deviations for variables by condition, revealing apparent differences between the conditions.

\begin{table*}[ht]
\centering
\caption{Variable Means, Standard Deviations, and 95\% Confidence Intervals by Condition}
\label{tab:means_sds}
\begin{tabular}{lccc}
\toprule
Variable
  & Low
  & Medium
  & High \\
\midrule
Intelligence
  & 3.88 $\pm$ 0.91 [3.617, 4.135]
  & 4.31 $\pm$ 0.90 [4.057, 4.567]
  & 3.82 $\pm$ 1.11 [3.504, 4.136] \\
Enjoyment
  & 4.93 $\pm$ 1.54 [4.496, 5.370]
  & 5.67 $\pm$ 1.24 [5.320, 6.026]
  & 5.43 $\pm$ 1.29 [5.068, 5.799] \\
Anthropomorphism
  & 2.82 $\pm$ 1.01 [2.536, 3.112]
  & 3.48 $\pm$ 1.02 [3.189, 3.771]
  & 3.06 $\pm$ 1.01 [2.778, 3.350] \\
Intention to Adopt
  & 4.31 $\pm$ 1.62 [3.851, 4.769]
  & 5.26 $\pm$ 1.35 [4.878, 5.642]
  & 4.58 $\pm$ 1.58 [4.130, 5.030] \\
Trust
  & 5.30 $\pm$ 0.96 [5.025, 5.571]
  & 5.79 $\pm$ 0.96 [5.520, 6.064]
  & 5.52 $\pm$ 1.02 [5.231, 5.813] \\
Likeability
  & 3.81 $\pm$ 0.95 [3.537, 4.079]
  & 4.47 $\pm$ 0.72 [4.268, 4.676]
  & 3.94 $\pm$ 1.05 [3.639, 4.233] \\
\bottomrule
\end{tabular}
\end{table*}

Kruskal--Wallis tests revealed significant effects for all six variables (see Table \ref{tab:kw_conditions}): Intelligence ($H(2) = 9.32$, $p = .010$, $\eta^2 = .049$), Enjoyment ($H(2) = 7.70$, $p = .021$, $\eta^2 = .038$), Anthropomorphism ($H(2) = 10.81$, $p = .005$, $\eta^2 = .059$), Intention to Adopt ($H(2) = 11.02$, $p = .004$, $\eta^2 = .061$), Trust ($H(2) = 8.39$, $p = .015$, $\eta^2 = .043$), and Likeability ($H(2) = 16.64$, $p < .001$, $\eta^2 = .098$).

\begin{table}[t]
\centering
\caption{Kruskal--Wallis tests for variables across three conditions}
\label{tab:kw_conditions}
\begin{tabular}{@{}lrrcr@{}}
\toprule
Variable & $H$ & $p$& $\eta^{2}$ & Sig.\\
\midrule
Intelligence                     &  9.317 & .010   & 0.049 & ** \\
Enjoyment                        &  7.703 & .021   & 0.038 & *  \\
Anthropomorphism                 & 10.812 & .005   & 0.059 & ** \\
Intention to Adopt               & 11.019 & .004   & 0.061 & ** \\
Trust                            &  8.393 & .015   & 0.043 & *  \\
Likeability                      & 16.635 & $<.001$& 0.098 & *** \\
\bottomrule
\end{tabular}
\vspace{2mm}
\begin{flushleft}
\footnotesize \textit{Note.} Significance codes: $^{*}p<.05$, $^{**}p<.01$, $^{***}p<.001$. $\eta^{2}$ denotes rank-based effect size.
\end{flushleft}
\end{table}

\begin{figure*}[ht]
\centering
\includegraphics[width=\textwidth]{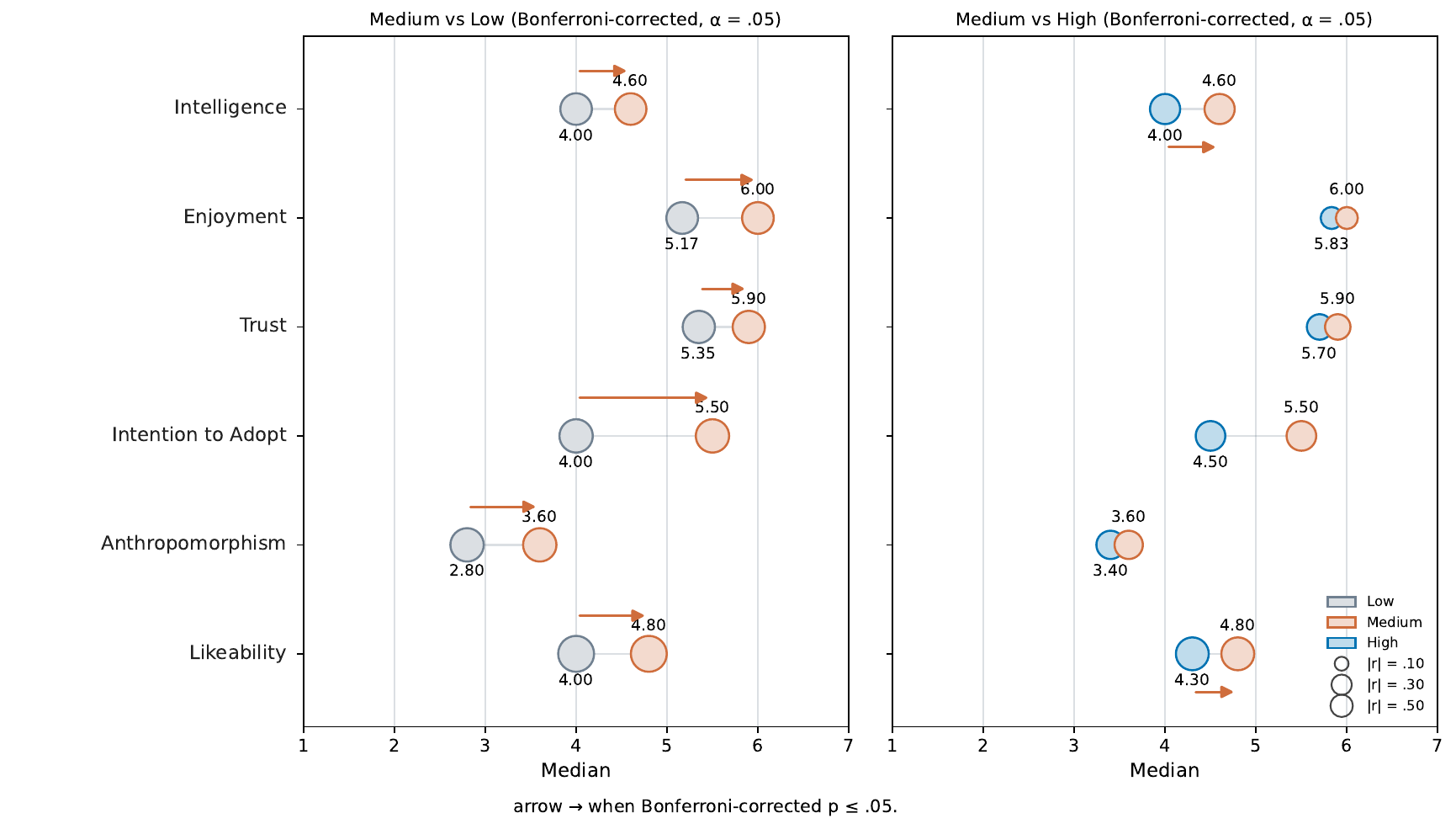}
\caption{Post hoc Mann--Whitney $U$ comparisons with Bonferroni adjustment ($\alpha = .05$). The left panel shows Medium vs Low CA, and the right panel shows Medium vs High CA across six outcomes. Circles mark group medians (colored by condition) and circle size reflects $|r|$ (.10/.30/.50). Arrows indicate comparisons that remain significant after correction ($p_{\text{Bonf}} \le .05$) and point toward the higher median. Medium CA exceeds Low CA on all six outcomes; relative to High CA, Medium CA is higher only for Likeability and Intelligence.}
\Description{Two side-by-side panels titled “Medium vs Low” and “Medium vs High.” The y-axis lists outcomes (Intelligence, Perceived Enjoyment, Trust, Perceived Intention to Adopt, Anthropomorphism, Likeability) and the x-axis shows the median scale. In the left panel, gray circles (Low) and orange circles (Medium) are plotted for each outcome, with orange right-pointing arrows indicating significant Medium superiority; median pairs (Low $\rightarrow$ Medium) are: Intelligence 4.00 $\rightarrow$ 4.60; Enjoyment 5.17 $\rightarrow$ 6.00; Trust 5.35 $\rightarrow$ 5.90; Intention to Adopt 4.00 $\rightarrow$ 5.50; Anthropomorphism 2.80 $\rightarrow$ 3.60; Likeability 4.00 $\rightarrow$ 4.80. In the right panel, blue circles (High) and orange circles (Medium) are plotted; arrows appear only for Intelligence (4.00 $\rightarrow$ 4.60) and Likeability (4.30 $\rightarrow$ 4.80), indicating significant Medium $>$ High. No arrows are shown for Enjoyment (5.83 vs.\ 6.00), Trust (5.70 vs.\ 5.90), or Intention to Adopt (4.50 vs.\ 5.50), denoting non-significant differences. A legend indicates colors for Low/Medium/High and circle sizes for $|r|{=}.10, .30, .50$.}
\label{fig:posthoc_condition}
\end{figure*}

Bonferroni-adjusted Mann--Whitney $U$ tests ($\alpha = .017$) showed that the medium condition consistently outperformed the low condition on all identified above (see Figure \ref{fig:posthoc_condition}, Table \ref{tab:posthoc_mwu_levels}). For Intelligence (medium: 4.60 vs. low: 4.00), $U = 850.0$, $p = .016$, $r = .276$; the medium condition also outperformed the high condition (medium: 4.60 vs. high: 4.00), $U = 1605.5$, $p = .039$, $r = .245$. Similar patterns were observed for Enjoyment (medium: 6.00 vs. low: 5.17), $U = 847.0$, $p = .015$, $r = .278$; and for Anthropomorphism (medium: 3.60 vs. low: 2.80), $U = 789.5$, $p = .005$, $r = .317$. Intention to Adopt likewise favored the medium condition (medium: 5.50 vs. low: 4.00), $U = 792.0$, $p = .005$, $r = .316$. For Trust, medium ($\mathit{Mdn} = 5.90$) exceeded low ($\mathit{Mdn} = 5.35$), $U = 826.5$, $p = .011$, $r = .292$. Finally, for Likeability (medium: 4.80 vs. low: 4.00), $U = 700.5$, $p < .001$, $r = .379$; the medium condition also outperformed the high condition (medium: 4.80 vs. high: 4.30), $U = 1692.0$, $p = .006$, $r = .305$.

\begin{table*}[t]
\centering
\caption{Post hoc Mann--Whitney $U$ tests (Bonferroni correction) for pairwise comparisons across three levels}
\label{tab:posthoc_mwu_levels}
\begin{tabular}{@{}llrrcrr@{}}
\toprule
Variable & Pair & $U$ & $p$ & $p_{\mathrm{adj}}$ & $r$ & Sig. \\
\midrule
Intelligence & low vs medium  & 850.0  & .005 & .016 & 0.276 & * \\
& low vs high    & 1238.0 & .936 & 1.000 & 0.008 & \\
& medium vs high & 1605.5 & .013 & .039 & 0.245 & * \\
\addlinespace
Enjoyment & low vs medium  & 847.0  & .005 & .015 & 0.278 & * \\
& low vs high    & 1010.5 & .098 & .293 & 0.165 & \\
& medium vs high & 1387.5 & .338 & 1.000 & 0.095 & \\
\addlinespace
Anthropomorphism & low vs medium  & 789.5  & .002 & .005 & 0.317 & ** \\
& low vs high    & 1057.5 & .185 & .554 & 0.133 & \\
& medium vs high & 1550.0 & .038 & .115 & 0.207 & \\
\addlinespace
Intention to Adopt & low vs medium  & 792.0  & .002 & .005 & 0.316 & ** \\
& low vs high    & 1119.0 & .365 & 1.000 & 0.090 & \\
& medium vs high & 1589.5 & .019 & .056 & 0.234 & \\
\addlinespace
Trust & low vs medium  & 826.5  & .004 & .011 & 0.292 & * \\
& low vs high    & 1063.0 & .198 & .594 & 0.129 & \\
& medium vs high & 1472.0 & .126 & .379 & 0.153 & \\
\addlinespace
Likeability & low vs medium  & 700.5  & $<.001$ & $<.001$ & 0.379 & *** \\
& low vs high    & 1115.5 & .353 & 1.000 & 0.093 & \\
& medium vs high & 1692.0 & .002 & .006 & 0.305 & ** \\
\bottomrule
\end{tabular}
\vspace{2mm}
\begin{flushleft}
\footnotesize \textit{Note.} Pairwise comparisons for variables with significant Kruskal--Wallis results. $p_{\mathrm{adj}}$ values are Bonferroni-corrected. Significance codes: $^{*}p_{\mathrm{adj}}<.05$, $^{**}p_{\mathrm{adj}}<.01$, $^{***}p_{\mathrm{adj}}<.001$. $U$ = Mann--Whitney statistic; $r$ = rank-biserial effect size.
\end{flushleft}
\end{table*}

\begin{framed}
\noindent\textbf{\textit{Summary.}} Medium CAs consistently produced significantly higher perceived Intelligence, Enjoyment, Anthropomorphism, Intention to Adopt, Trust, and Likeability than Low CAs, and also yielded significantly greater perceived Intelligence and Likeability than High CAs. This pattern suggests an inverted-U relationship between personality expression level and user experience.
\end{framed}

\subsection{Personality Alignment and User Perceptions (RQ2)}
\begin{figure}[ht]
  \centering
  \includegraphics[width=\columnwidth]{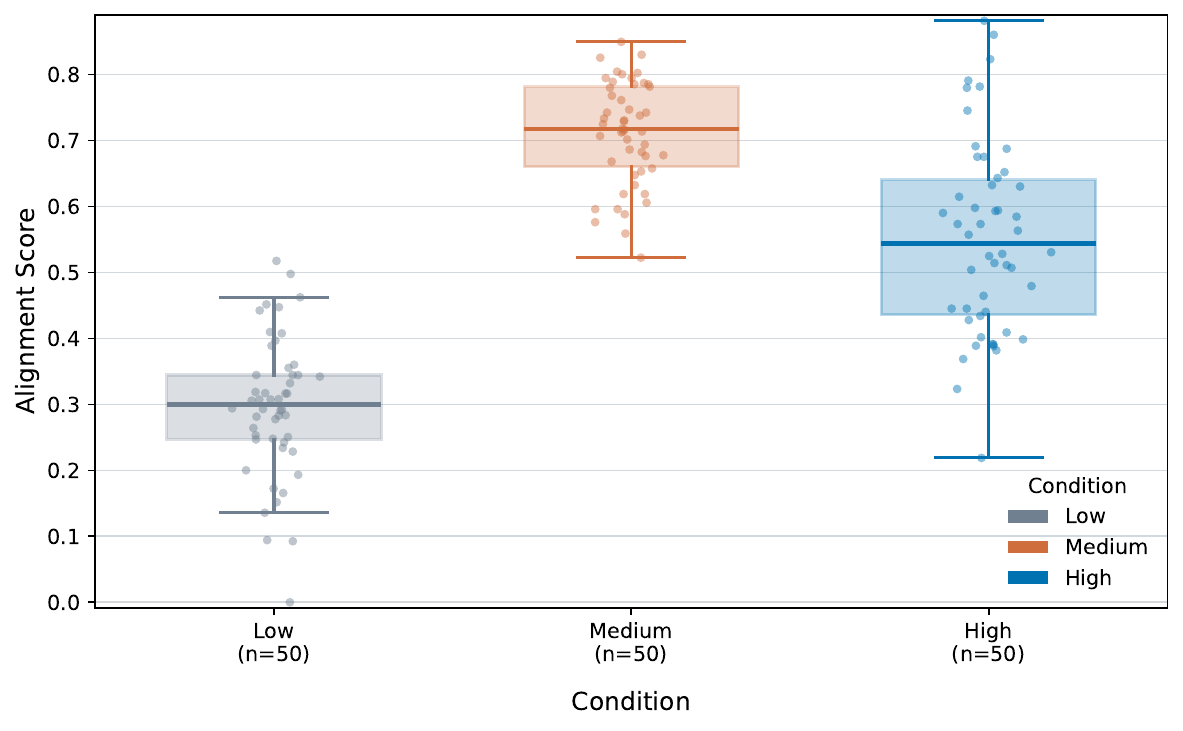}
  \caption{Distribution of user–CA Personality Alignment Scores (0–1) by experimental condition ($n=50$ per group). Dots show participants; boxplots show medians and interquartile ranges. The medium condition exhibits the highest and tightest alignment (median $\approx .72$), the high condition shows a wider midrange spread (median $\approx .54$), and the low condition has the lowest alignment (median $\approx .30$).}
  \Description{Three color-coded boxplots with jittered dots on a shared y-axis labeled “Alignment Score” ranging from 0 to approximately 0.85. Left: a gray boxplot for Low ($n=50$) with dots clustered around 0.25--0.35 and a median near 0.30. Center: an orange boxplot for Medium ($n=50$) with dots concentrated around 0.65--0.80 and a median near 0.72. Right: a blue boxplot for High ($n=50$) with a broad spread, dots ranging from roughly 0.22 to about 0.86, and a median around 0.54. A legend at the right labels the Low, Medium, and High conditions.}
\label{fig:alignment_score}
\end{figure}

Personality alignment scores between users and CAs ($M = 0.52$, $SD = 0.21$, range $= 0$--$0.88$) showed consistent positive associations between personality alignment and user perceptions. As illustrated in Figure \ref{fig:alignment_score}, the distribution of alignment scores exhibits substantial within-condition variability that enabled examining alignment effects beyond experimental manipulations.

\begin{table}[ht]
\centering
\caption{Spearman correlations with personality alignment score; *** $p \le .001$, ** $p \le .01$, * $p \le .05$}
\label{tab:spearman-correlations}
\begin{tabular}{lrc}
\toprule
Variable & $r_s$ & Sig. \\
\midrule
Intelligence                      & .253 & **  \\
Enjoyment                         & .234 & **  \\
Anthropomorphism                  & .231 & **  \\
Intention to Adopt                & .269 & *** \\
Trust                             & .288 & *** \\
Likeability                       & .304 & *** \\
\bottomrule
\end{tabular}
\end{table}

Spearman's rank correlations indicated positive relationships between personality alignment and all variables, as shown in Table \ref{tab:spearman-correlations}. Likeability ($r_s = .304$, $p < .001$) was strongest, followed by Trust ($r_s = .288$, $p < .001$), Intention to Adopt ($r_s = .269$, $p < .001$), Intelligence ($r_s = .253$, $p = .002$), Enjoyment ($r_s = .234$, $p = .004$), and Anthropomorphism ($r_s = .231$, $p = .005$).

While personality alignment showed significant bivariate correlations with user perceptions, hierarchical regression analyses controlling for experimental condition revealed more limited effects. Trust showed a significant incremental contribution from personality alignment ($\beta = 1.397$, $p = .084$; $\Delta R^2 = .025$, $p = .049$), and Intelligence demonstrated a similar pattern ($\beta = 1.185$, $p = .076$; $\Delta R^2 = .018$, $p = .094$). Enjoyment ($\beta = 1.347$, $p = .196$; $\Delta R^2 = .012$, $p = .172$), Intention to Adopt ($\beta = 1.728$, $p = .156$; $\Delta R^2 = .016$, $p = .116$), and Likeability ($\beta = 0.790$, $p = .188$; $\Delta R^2 = .009$, $p = .234$) showed consistent directional trends in the expected direction. These effect sizes are practically meaningful, with alignment improvements translating to substantial changes in user perceptions. However, the current sample size constrains statistical power. These findings underscore the importance of personality alignment but suggest that larger samples are necessary to reliably detect incremental effects.

\begin{framed}
\noindent\textbf{\textit{Summary.}} Personality alignment demonstrated consistent positive correlations with all six variables, suggesting that beyond the main effects of CA personality design, individual user–agent personality alignment contributes meaningfully to user experience.
\end{framed}

\subsection{Individual Trait Distance Effects (RQ3)}
Analysis of individual trait distances revealed differential impacts on user perceptions. Table \ref{tab:trait-correlations} reports Spearman correlations between each trait mismatch and the variables. Conscientiousness distance showed the strongest and most pervasive negative associations, correlating with Intelligence ($r_s = -.231$, $p = .005$) and Likeability ($r_s = -.288$, $p < .001$); its association with Trust was non-significant ($r_s = -.146$, $p = .075$). Extraversion distance was broadly impactful, relating negatively to  Intelligence ($r_s = -.232$, $p = .004$), Intention to Adopt ($r_s = -.227$, $p = .005$), Trust ($r_s = -.202$, $p = .013$), and Likeability ($r_s = -.271$, $p = .001$). Emotional Stability distance influenced evaluative judgments like Intelligence ($r_s = -.222$, $p = .006$), Intention to Adopt ($r_s = -.166$, $p = .042$), Trust ($r_s = -.191$, $p = .020$), and Likeability ($r_s = -.233$, $p = .004$). Agreeableness distance impacted Enjoyment ($r_s = -.208$, $p = .011$), Intention to Adopt ($r_s = -.203$, $p = .013$), and Trust ($r_s = -.188$, $p = .021$). Openness distance showed significant associations with two variables: Intention to Adopt ($r_s = -.164$, $p = .045$) and Trust ($r_s = -.201$, $p = .014$), while showing non-significant correlations with the remaining four variables. The consistent negative direction of these correlations indicates that greater mismatches between user and CA personality traits are associated with lower values of the variables.

\begin{table*}[ht]
  \caption{Spearman correlations between trait distances and variables}
  \label{tab:trait-correlations}
  \begin{tabular}{@{}llrrr@{}}
    \toprule
    Variable & Trait Distance & $r_s$ & $p$ & Sig. \\
    \midrule
    Intelligence       & Openness            & -.071 & .386  &     \\
                                & Conscientiousness   & -.231 & .005  & **  \\
                                & Extraversion        & -.232 & .004  & **  \\
                                & Agreeableness       & -.114 & .166  &     \\
                                & Emotional Stability & -.222 & .006  & **  \\
    \midrule
    Enjoyment         & Openness            & -.143 & .081  &     \\
                                & Conscientiousness   & -.146 & .075  &     \\
                                & Extraversion        & -.147 & .073  &     \\
                                & Agreeableness       & -.208 & .011  & *   \\
                                & Emotional Stability & -.117 & .153  &     \\
    \midrule
    Anthropomorphism   & Openness            & -.078 & .341  &     \\
                                & Conscientiousness   & -.134 & .101  &     \\
                                & Extraversion        & -.163 & .046  & *   \\
                                & Agreeableness       & -.133 & .105  &     \\
                                & Emotional Stability & -.131 & .111  &     \\
    \midrule
    Intention to Adopt 
                                & Openness            & -.164 & .045  & *   \\
                                & Conscientiousness   & -.090 & .273  &     \\
                                & Extraversion        & -.227 & .005  & **  \\
                                & Agreeableness       & -.203 & .013  & *   \\
                                & Emotional Stability & -.166 & .042  & *   \\
    \midrule
    Trust              & Openness            & -.201 & .014  & *   \\
                                & Conscientiousness   & -.146 & .075  &     \\
                                & Extraversion        & -.202 & .013  & *   \\
                                & Agreeableness       & -.188 & .021  & *   \\
                                & Emotional Stability & -.191 & .020  & *   \\
    \midrule
    Likeability        & Openness            & -.126 & .123  &     \\
                                & Conscientiousness   & -.288 & $< .001$ & *** \\
                                & Extraversion        & -.271 & $< .001$ & *** \\
                                & Agreeableness       & -.121 & .139  &     \\
                                & Emotional Stability & -.233 & .004  & **  \\
    \bottomrule
  \end{tabular}
  \begin{flushleft}
    \textbf{Note.} * $p < .05$, ** $p < .01$, *** $p < .001$. Trait distance variables reflect the absolute difference between participant and CA scores on each trait. Negative correlations indicate that greater trait distance is associated with lower values of the variables.
  \end{flushleft}
\end{table*}

\begin{framed}
\noindent\textbf{\textit{Summary.}} Individual trait distances showed differential effects. Conscientiousness mismatches most negatively affected Intelligence and Likeability. Extraversion mismatches were broadly impactful, relating to Intelligence, Anthropomorphism, Intention to Adopt, Trust, and Likeability. Emotional Stability mismatches reduced Intelligence, Intention to Adopt, Trust, and Likeability. Agreeableness mismatches lowered Enjoyment, Intention to Adopt, and Trust. Openness mismatches had comparatively minimal impact, with small effects on Intention to Adopt and Trust.
\end{framed}

\subsection{Personality Alignment Clusters (RQ4)}
k-means clustering based on trait mismatch patterns identified three distinct user groups, jointly supported by silhouette analysis (score $= .310$) and elbow-curve  (see Appendix \ref{appendix:analyses} for details). Figure \ref{fig:alignment_cluster} depicts the resulting profiles.

\begin{figure}[ht]
  \centering
  \includegraphics[width=\columnwidth]{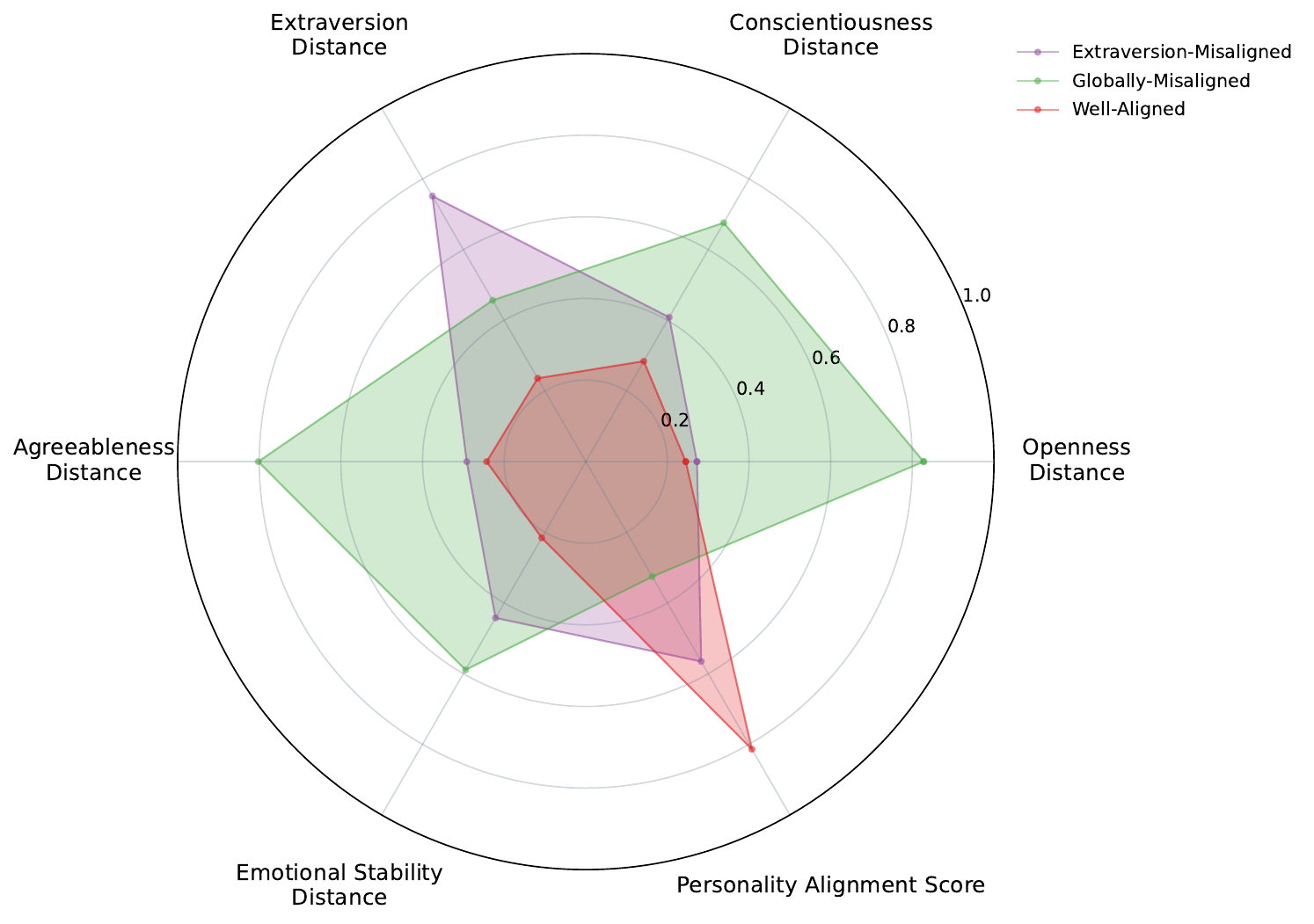}
  \caption{Personality alignment clusters from $k$-means on trait distances (Openness, Conscientiousness, Extraversion, Agreeableness, Emotional Stability; 0–1) plus overall Personality Alignment Score. The radar plot shows three profiles: Extraversion-Misaligned (purple) has a large Extraversion distance with otherwise moderate distances and a mid alignment score; Globally-Misaligned (green) shows high distances on all traits and the lowest alignment; Well-Aligned (red) exhibits uniformly small distances and the highest alignment score.}
  \Description{Radar chart with six spokes labeled Openness Distance, Conscientiousness Distance, Extraversion Distance, Agreeableness Distance, Emotional Stability Distance, and Personality Alignment Score, with radial scale marks at 0.2, 0.4, 0.6, 0.8, and 1.0. Three filled polygons are shown: a purple “Extraversion-Misaligned” cluster that spikes near 0.75 on Extraversion Distance and sits around 0.25--0.45 on the other trait distances, with an overall alignment of approximately 0.5; a green “Globally-Misaligned” cluster that is large on most spokes (Openness $\approx$ 0.8, Conscientiousness $\approx$ 0.7, Agreeableness $\approx$ 0.8, Emotional Stability $\approx$ 0.6, Extraversion $\approx$ 0.5) and has low alignment ($\approx$ 0.3); and a red “Well-Aligned” cluster that remains small on all trait distances ($\approx$ 0.2--0.3) and high on alignment ($\approx$ 0.8). A legend labels the three clusters.}
\label{fig:alignment_cluster}
\end{figure}

Cluster 0 ($n = 39$, 26.0\%) represented ``Extraversion-Misaligned'' users, characterized by high extraversion distance mismatches ($M = 3.01$, $SD = 0.67$, 95\% CI [2.790, 3.223]) alongside comparatively smaller mismatches on other traits (e.g., Agreeableness: $M = 1.17$, $SD = 0.75$, 95\% CI [0.923, 1.411]). This cluster consisted primarily of high-condition participants (92.3\%) and showed a moderate overall alignment score ($M = 0.498$, $SD = 0.090$, 95\% CI [0.469, 0.528]). Cluster 1 ($n = 49$, 32.7\%) comprised ``Globally-Misaligned'' users showing high trait distance across multiple traits. This cluster contained almost exclusively low-condition participants (98.0\%) and exhibited the lowest personality alignment ($M = 0.287$, $SD = 0.099$, 95\% CI [0.259, 0.315]). Cluster 2 ($n = 62$, 41.3\%) included ``Well-Aligned'' users with the highest personality alignment scores ($M = 0.717$, $SD = 0.083$, 95\% CI [0.696, 0.739]). This cluster predominantly featured medium-condition participants (77.4\%).

\begin{figure*}[ht]
\centering
\includegraphics[width=\textwidth]{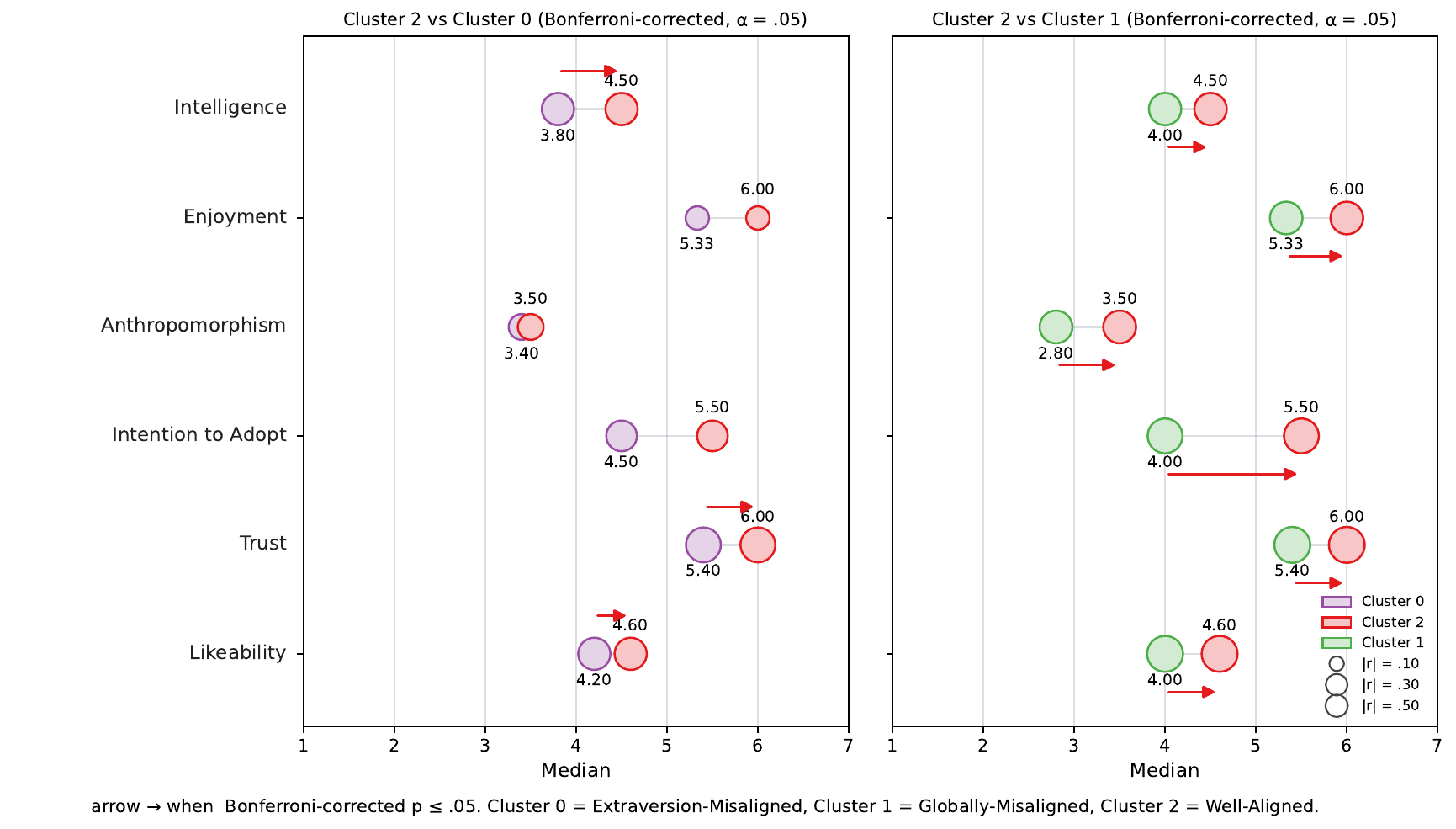}
\caption{Post hoc Mann--Whitney $U$ comparisons with Bonferroni adjustment ($\alpha = .05$). The left panel shows Cluster 2 (Well-Aligned) vs Cluster 0 (Extraversion-Misaligned), and the right panel shows Cluster 2 vs Cluster 1 (Globally-Misaligned) across all outcomes. Circles mark cluster medians (colored by cluster), with size reflecting $|r|$ (.10/.30/.50). Arrows indicate comparisons that remain significant after correction ($p_{\text{Bonf}} \le .05$) and point toward the higher median. Cluster 2 exceeds Cluster 0 on Intelligence, Trust, and Likeability. Cluster 2 exceeds Cluster 1 on all six outcomes.}
\Description{Two panels of median-dot plots comparing outcome medians across alignment clusters. Left panel, titled “Cluster 2 vs Cluster 0,” shows purple dots (C0) and red dots (C2) for each outcome along an x-axis labeled “Median.” Red right-pointing arrows denote significant C2 superiority for Intelligence (C0 3.80, C2 4.50), Trust (5.40 $\rightarrow$ 6.00), and Likeability (4.20 $\rightarrow$ 4.60). No arrows are shown for Perceived Enjoyment (5.33 vs.\ 6.00), Intention to Adopt (4.50 vs.\ 5.50), or Anthropomorphism (3.40 vs.\ 3.50), indicating non-significant differences. Right panel, titled “Cluster 2 vs Cluster 1,” shows green dots (C1) and red dots (C2), with red arrows indicating significant C2 superiority for all six outcomes: Intelligence (4.00 $\rightarrow$ 4.50), Perceived Enjoyment (5.33 $\rightarrow$ 6.00), Anthropomorphism (2.80 $\rightarrow$ 3.50), Intention to Adopt (4.00 $\rightarrow$ 5.50), Trust (5.40 $\rightarrow$ 6.00), and Likeability (4.00 $\rightarrow$ 4.60). A legend indicates colors for C0, C1, and C2 and circle sizes for $|r|{=}.10, .30, .50$.}
\label{fig:posthoc_cluster}
\end{figure*}

Kruskal--Wallis tests in Table \ref{tab:kw_clusters} indicate significant differentiation across all six variables (all $p < .05$; see also Figure \ref{fig:posthoc_cluster}), suggesting that personality alignment patterns are systematically associated with user perceptions. Post hoc Mann--Whitney $U$ comparisons (Table \ref{tab:posthoc_mwu_clusters}) show that ``Well-Aligned'' users (Cluster 2) consistently report higher values than the remaining clusters on all significant variables. The largest effects emerge for Enjoyment (Cluster 2 vs.\ Cluster 1: $U = 1062.5$, $p = .006$, $r = .257$) and Trust (Cluster 2 vs.\ Cluster 1: $U = 945.5$, $p < .001$, $r = .323$). 

\begin{table}[t]
  \centering
  \caption{Cluster-based Kruskal--Wallis tests for variables across three personality alignment clusters}
  \label{tab:kw_clusters}
  \begin{tabular}{@{}lrrrc@{}}
    \toprule
    Variable & $H$ & $p$ & $\eta^{2}$ & Sig. \\
    \midrule
    Intelligence                     &  9.650 & .008     & .051 & **  \\
    Enjoyment                        &  6.987 & .030     & .033 &  *  \\
    Anthropomorphism                 &  8.037 & .018     & .041 &  *  \\
    Intention to Adopt               & 11.340 & .003     & .063 & **  \\
    Trust                            & 14.891 & $< .001$ & .087 & *** \\
    Likeability                      & 13.357 & .001     & .076 & **  \\
    \bottomrule
  \end{tabular}
  \vspace{2mm}
  \begin{flushleft}
  \footnotesize \textit{Note.} Significance codes: $^{*}p<.05$, $^{**}p<.01$, $^{***}p<.001$. $\eta^{2}$ denotes rank-based effect size.
  \end{flushleft}
\end{table}

\begin{table*}[t]
  \centering
  \caption{Post hoc Mann--Whitney $U$ tests for clusters (Bonferroni correction)}
  \label{tab:posthoc_mwu_clusters}
  \begin{tabular}{@{}llrrcrr@{}}
    \toprule
    Construct & Pair & $U$ & $p$ & $p_{\mathrm{adj}}$ & $r$ & Sig. \\
    \midrule
    Intelligence & C0 vs C1 & 916.5  & .744 & 1.000 & .035 & \\
                 & C0 vs C2 & 851.0  & .011 & .034  & .248 & * \\
                 & C1 vs C2 & 1074.5 & .008 & .023  & .251 & * \\
    \addlinespace
    Enjoyment & C0 vs C1 & 1090.5 & .256 & .768 & .121 & \\
              & C0 vs C2 & 1064.0 & .307 & .921 & .101 & \\
              & C1 vs C2 & 1062.5 & .006 & .019 & .257 & * \\
    \addlinespace
    Anthropomorphism & C0 vs C1 & 1128.5 & .146 & .439 & .155 & \\
                     & C0 vs C2 & 1013.0 & .172 & .515 & .136 & \\
                     & C1 vs C2 & 1059.5 & .006 & .019 & .259 & * \\
    \addlinespace
    Intention to Adopt & C0 vs C1 & 1054.5 & .405 & 1.000 & .089 & \\
                       & C0 vs C2 & 893.5  & .027 & .080  & .219 & \\
                       & C1 vs C2 & 983.0  & .001 & .004  & .302 & ** \\
    \addlinespace
    Trust & C0 vs C1 & 980.5  & .837 & 1.000 & .022 & \\
          & C0 vs C2 & 773.5  & .002 & .007  & .302 & ** \\
          & C1 vs C2 & 945.5  & $< .001$ & .002  & .323 & ** \\
    \addlinespace
    Likeability & C0 vs C1 & 1041.0 & .473 & 1.000 & .077 & \\
               & C0 vs C2 & 851.0  & .012 & .035  & .248 & * \\
               & C1 vs C2 & 944.5  & $< .001$ & .002  & .324 & ** \\
    \bottomrule
  \end{tabular}
  \vspace{2mm}
  \begin{flushleft}
  \footnotesize \textit{Note.}  Pairwise comparisons for variables with significant cluster-based Kruskal--Wallis results. $p_{\mathrm{adj}}$ values are Bonferroni-corrected. Significance codes: $^{*}p_{\mathrm{adj}}<.05$, $^{**}p_{\mathrm{adj}}<.01$. $U$ = Mann--Whitney statistic; $r$ = rank-biserial effect size. C0 = Extraversion-Misaligned, C1 = Globally-Misaligned, C2 = Well-Aligned.
  \end{flushleft}
\end{table*}

Within-cluster correlation analyses revealed differential sensitivity to alignment. ``Extraversion-Misaligned'' users (Cluster 0) showed positive correlations between personality alignment score and Intelligence ($r_s = .385$, $p = .016$) as well as Trust ($r_s = .412$, $p = .009$), suggesting compensatory effects when other traits align well despite Extraversion mismatches.

\begin{framed}
\noindent\textbf{\textit{Summary.}} Three distinct clusters emerged based on personality alignment patterns. The ``Well-Aligned'' cluster showed the highest values across variables. Post hoc tests (Bonferroni-corrected) showed that the ``Well-Aligned'' cluster's values for Trust, Likeability, and Intelligence were higher than both other clusters, and values for Enjoyment, Intention to Adopt, and Anthropomorphism were higher than the ``Globally-Misaligned'' cluster. The concentration of ``Well-Aligned'' users in the medium condition (77.4\%), relative to the high and low conditions, reinforces that moderate personality expression facilitates broader compatibility.
\end{framed}

\subsection{Qualitative Perceptions}
To complement the quantitative results, we also invited participants to leave qualitative opinions about their experience with the CA. 114 of 150 participants (76\%) provided some qualitative data. An inductive thematic analysis revealed three overarching themes:

\subsubsection{Medium CA Appeal}
Participants described the Medium CA as balanced, approachable, and effective. Comments highlighted friendliness (P1), attentiveness (P17), and a personality that ``meshed with mine'' (P24). Others emphasized trustworthiness, with a style ``fluid and to the point'' (P15). While a few noted a slightly “robotic” quality (P13), overall reflections reinforced higher Intelligence, Intention to Adopt, and Likeability.

\subsubsection{Polarizing Low and High CA}
The Low and High CAs elicited more divided responses. The Low CA was often described as flat or lacking presence: ``It didn't seem to have much of a personality'' (P51), ``a little flat'' (P53), or less enthusiastic than the user (P82). 

The High CA, by contrast, was seen as exaggerated and less authentic. Some found it ``great'' and relatable (P103), but many described it as grating: ``The personality was exhausting... like a teenage cheerleader... it detracts from the authenticity and credibility of the AI'' (P114). These reactions underscore why the high condition was judged less likable and intelligent than the medium, showing how an amplified persona can feel ``distracting'' (P121).

\subsubsection{Personality Alignment Effects}
Personality fit emerged as a critical factor even in qualitative accounts. Apparent mismatches produced the most negative reactions: for instance, one high-extra\-verted user criticized the low-extraversion agent as ``very robotic'' and out of step with their enthusiastic tone (P51), echoing the correlation patterns observed for RQ3.

Conversely, strong alignment elevated the interaction. The participant who said the agent's personality ``meshed with mine'' (P24) was in the medium condition and had a personality alignment score of $0.648$,  indicating a close match to their trait profile. These comments indicate that effective interaction depends not only on agent design in isolation but also on the specific user–agent fit.

\begin{framed}
\noindent\textbf{\textit{Summary.}} Qualitative analysis revealed three key themes: (1) Medium CA was perceived as balanced and professional, (2) Low CA lacked social presence while High CA appeared inauthentic, and (3) personality alignment emerged as critical, with mismatches generating negative reactions.
\end{framed}

\section{Discussion}

Our investigation reveals systematic patterns in how CA personality design shapes user perceptions, extending prior work on personality expression and user–agent compatibility in HCI.

A core theoretical contribution of this work lies in reconceptualizing personality alignment in human–AI interaction beyond the categorical binaries that have constrained prior research. Prior HCI studies have largely treated alignment as simple binary matching—users and agents either correspond on a trait or they do not \cite{isbister_consistency_2000, zhou_trusting_2019, volkel_user_2022, spagnolli_similarity_2025}. This framing oversimplifies what is fundamentally a continuous phenomenon. At the same time, current LLMs still resist fine-grained, continuous personality control \cite{ramirez_controlling_2023, jiang_evaluating_2023, jiang_personallm_2024}. Our hybrid approach resolves this tension: we implement personality through categorical bands (low, medium, high) while measuring alignment continuously using Euclidean distance between user and agent personality profiles. This design enables analysis of both overall condition effects and the within-condition individual differences that purely categorical approaches cannot capture.

This methodological advance further reveals that personality traits exert \textbf{differential importance} in determining user–agent compatibility. Instead of uniform effects across all Big Five traits, Extraversion and Emotional Stability emerge as especially consequential for positive user perceptions, whereas Openness has minimal impact in goal-oriented contexts. This pattern shifts the understanding of alignment from holistic matching to selective compatibility: certain traits disproportionately shape user experience. These trait-specific effects allow for more efficient design strategies that prioritize the dimensions that matter most, rather than attempting comprehensive alignment across all traits.

Together, these insights—the continuous nature of alignment and the uneven influence of specific traits—expose the profound \textbf{ethical dimensions} of personality design in artificial agents. Personality alignment effectively operates as a ‘‘trust slider’’ capable of modulating user experience, adoption intentions, and perceived agent credibility through mechanisms users may not consciously register. That these effects appear even in a constrained travel-planning scenario highlights their potential intensity in higher-stakes domains such as health or finance, where trust carries significant consequences. The risks are amplified for vulnerable populations—particularly older adults \cite{chin_being_2021, pradhan_use_2020, pradhan_phantom_2019}—who may be especially susceptible to such influence. Our findings, therefore, position personality as a \textbf{critical vector of technological influence}, one that requires ethical frameworks capable of addressing both its manipulative potential and its functional role in shaping user satisfaction and system usability.

Together, these insights set the stage for the sections that follow. We begin by examining why users systematically preferred medium-intensity personalities—a pattern that reveals a deeper divide between human-human and human-CA social heuristics. We then translate these findings into concrete design implications for building personality-aware systems that balance usability, alignment, and ethical constraints. Finally, we outline the study’s limitations and identify future research directions that can extend TMK and deepen understanding of CA personality effects across domains and contexts.

\subsection{The Human-CA Personality Divide}
The preference for the Medium CA observed in our study diverges sharply from established patterns in social psychology. For example, \citet{van_der_linden_general_2010} describe the general factor of personality (GFP) as a continuum from relatively undesirable to highly desirable profiles, with higher GFP scores reflecting higher Extraversion, Agreeableness, and Emotional Stability. Such profiles are typically associated with greater cooperation, support, and positive impressions in human–human encounters. Classic work likewise shows that highly expressive and extraverted individuals are evaluated more favorably in initial interactions \cite{friedman_nonverbal_1988}, and that expressive individuals appear more attractive irrespective of physical appearance \cite{depaulo_expressiveness_1992}. These intuitions rely on human agency, reciprocity, and real interpersonal stakes.

If participants applied these same heuristics to CAs, the High CA should have outperformed the Medium CA. The High CA aligns with the high-GFP pattern that predicts greater social desirability, suggesting it should have produced higher perceived Intelligence, Likeability, and Trust. Yet participants consistently preferred the Medium CA, indicating a fundamental ontological gap in how users evaluate artificial versus human personalities.

Prior CA research has largely assumed that “more personality” translates to “better experience,” examining whether increasing expressiveness improves user perceptions \cite{10937459, zhang_exploring_2025, volkel_user_2022, amin_kuhail_assessing_2024, ghandeharioun_towards_2019}. Our findings challenge this assumption, revealing a clear inverted-U relationship: both ends of the spectrum were problematic, but for different reasons. Low CA was perceived as mechanically flat and lacking social presence. High CA, by contrast, raised authenticity concerns—participants questioned its sincerity and competence, which undermined perceived intelligence and trust. The Medium CA uniquely achieved the balance users sought: socially present yet professionally grounded, especially for users whose own personality profiles aligned with this level.

In other words, while strong personality intensity may carry social value in human–human contexts, in goal-oriented CAs it produces a performative penalty. This aligns with \citet{volkel_examining_2021}, who argue that CA design benefits from “moderate instead of extreme chatbot personalities,” marking the point where human social intuitions no longer transfer cleanly to artificial agents.

This preference for the Medium CA contrasts sharply with contemporary commercial LLM-based CAs, which increasingly default to high-intensity personalities. Non-reasoning LLMs such as GPT-4o, Gemini Flash 2.5, Deepseek, and Claude 3.7 Sonnet consistently exhibit ceiling effects in Agreeableness and Openness (Appendix Table \ref{tab:llm_personality}). Recent work shows that LLMs display a strong “social desirability bias” \cite{salecha_large_2024}, skewing their behavior toward hyper-desirable traits, and often adopt excessive agreeableness or flattery—forms of sycophancy that compromise judgment \cite{hu_quantifying_2024}. Industry practices, therefore, push agents toward precisely the High CA patterns our results identify as counterproductive, favoring performative friendliness over the functional moderation that users actually prefer in goal-oriented contexts.

\subsection{Design Implications}

The theoretical insights and empirical patterns described above converge on practical guidance for CA personality implementation, suggesting design implications (DIs) that move beyond fixed personality archetypes toward adaptive systems capable of accommodating the systematic variations in user-agent compatibility revealed by our investigation. These implications address how personality-aware systems can be designed to optimize user experience while maintaining ethical considerations, offering concrete strategies that build on our empirical findings about personality expression levels, trait-specific effects, and alignment patterns.

Medium CA provides the optimal design baseline, outperforming Low CA across all six user perception measures: Intelligence, Enjoyment, Anthropomorphism, Intention to Adopt, Trust, and Likeability. Medium CA also exceeded High CA on perceived Intelligence and Likeability, suggesting that overly intense personalities trigger authenticity concerns that undermine perceived competence and social acceptance; accordingly, Medium represents the safest default when designers prioritize trust formation and adoption intentions, as it maximizes positive user perceptions without inviting skepticism about agent credibility. Yet the existence of distinct alignment patterns between users and CA personalities demands adaptive calibration---systems that adapt personality expression beyond population-level optima to accommodate individual compatibility profiles. Effective calibration requires dual pathways: (1) user-facing mechanisms that enable direct personality adjustment through lightweight feedback (e.g., evaluating agent responses as ``too extraverted'' or ``overly agreeable,'' or selecting preferences from curated personality styles), and (2) system-driven approaches that allow CAs to develop ``mutual theory of mind'' \cite{wang_towards_2021} by inferring compatibility from interaction patterns, response engagement, and behavioral feedback over time. The design logic progresses from empirically grounded defaults through recognition of alignment heterogeneity to calibration mechanisms that balance user control with system learning, maintaining the transparency essential to ethical personality-aware AI systems.

A practical adaptive calibration mechanism can be implemented by combining lightweight onboarding prompts with interaction-level signals such as user edits, preference selections, or clarification requests. These signals can be mapped to TMK levels (low, medium, high) to adjust the CA’s personality intensity over time while maintaining stability. Such adaptation can be empirically evaluated by comparing user satisfaction, alignment scores, and task efficiency before and after calibration.

\begin{quote}
    
\textbf{DI1: Adaptive Calibration}. Begin with medium personality expression as an empirically grounded default, but implement adaptive mechanisms that enable both user-driven and system-driven personality adjustment to accommodate individual alignment patterns.

\end{quote}

Building on this foundation of adaptive calibration, our findings reveal differential trait effects that challenge uniform approaches to personality alignment, pointing toward trait-specific prioritization as a more efficient design strategy. The systematic variations in trait importance—Conscientiousness most strongly influencing Intelligence and Likeability; Extraversion broadly affecting Intelligence, Anthropomorphism, Intention to Adopt, Trust, and Likeability; Emotional Stability reducing Intelligence, Intention to Adopt, Trust, and Likeability; Agreeableness lowering Enjoyment, Intention to Adopt, and Trust; and Openness showing minimal impact overall—suggest that personality alignment should not be treated as a uniform process across all Big Five dimensions.

In resource-constrained design contexts, this prioritization approach offers significant advantages over comprehensive personality matching. Rather than attempting full-spectrum alignment across all traits, systems can achieve greater impact by focusing design energy on Extraversion and Emotional Stability, the dimensions most consequential for user experience in goal-oriented interactions. This selective personalization should incorporate contextual sensitivity, weighting traits differently depending on the task domain. Current industry practices already reflect this intuitive prioritization---OpenAI's recent updates\footnote{https://x.com/OpenAI/status/1956461718097494196} (announced on August 15th, 2025) emphasize making models more ``friendly and warm,'' targeting broad personality dimensions that align with user preferences. However, these efforts rely on impressionistic categories rather than the systematically defined trait dimensions established in personality psychology and psychometrics.

\begin{quote}
\textbf{DI2: Trait-Specific Prioritization}. Focus personality design resources on Extraversion and Emotional Stability rather than attempting comprehensive Big Five alignment, leveraging psychometrically validated trait dimensions to ground personality adjustments in systematic psychological frameworks.
\end{quote}

These calibration and prioritization strategies point toward a fundamental reconceptualization of personality within system design. Our findings demonstrate that personality expression produces measurable effects on perceived Intelligence and Likeability even within bounded, goal-oriented interactions, establishing personality as a functional UX lever rather than an incidental design flourish. These systematic effects across core user-experience outcomes challenge approaches that treat personality as cosmetic branding, and instead require that it be designed and evaluated alongside traditional UX dimensions such as accuracy, efficiency, and usability.

This functional perspective demands practical shifts in design and evaluation practices. Personality effects should be measured explicitly in UX testing rather than discovered incidentally, with systematic assessment of how expression shapes trust, comfort, and adoption intentions. Design decisions must balance task-oriented and social goals—expressive elements that enhance rapport cannot undermine perceptions of competence—while integrating personality into onboarding experiences so users can understand and shape interaction styles before deeper engagement. Such scaffolding reduces mismatched expectations and strengthens early trust formation.

Current industry practices often frame personality largely as brand voice, emphasizing qualities such as ``helpful, harmless, honest'' for Anthropic Claude\footnote{https://www.anthropic.com/news/introducing-claude}
 or Amazon’s portrayal of Alexa\footnote{https://developer.amazon.com/en-US/alexa/branding/alexa-guidelines/communication-guidelines/brand-voice}
 as ``clever, relevant, and making the customer smile,'' leaving users to discover agent personality implicitly through interaction. This approach reflects traditional marketing frameworks for brand personality \cite{Aaker_1997} rather than the systematic personality psychology that underlies effective human–computer social interaction \cite{nass_does_2001}. A more rigorous approach would surface personality as a deliberate component of onboarding and interaction design, treating it as a primary UX consideration that shapes expectations and adoption from initial contact. This shift from cosmetic to functional personality design recognizes that user–agent interpersonal fit fundamentally influences system success, warranting the same systematic attention as other usability dimensions.

Operationalizing personality as a UX parameter involves surfacing personality intensity as a configurable system control, similar to existing settings for verbosity, formality, or autonomy. Designers can offer preset styles mapped to TMK levels (e.g., concise, warm, structured), provide user-visible tone controls, and apply context-specific presets during different stages of a task. These practices make personality an adjustable UX lever rather than a fixed attribute. Such configurations can be evaluated through A/B tests comparing TMK-based presets and by measuring impacts on trust, cognitive load, satisfaction, and task progress across repeated interactions.

\begin{quote}
\textbf{DI3: Personality as Functional UX}. Treat personality as a core UX dimension that systematically influences user trust and adoption rather than cosmetic branding, integrating personality considerations into UX testing, design evaluation, and onboarding experiences with the same rigor applied to traditional usability concerns.
\end{quote}

This functional understanding of personality as a UX lever immediately raises profound ethical concerns that demand systematic safeguards rather than ad hoc protections. If personality alignment can systematically influence Intention to Adopt, Trust, and Likeability within bounded travel-planning tasks, these effects will likely intensify in high-stakes domains where users depend on AI systems for health decisions, financial guidance, or caregiving support. The ``trust slider'' capacity we observed represents a form of technological influence that could easily cross from user benefit into manipulation, particularly when applied to vulnerable populations or consequential decision contexts. 

Ethical safeguards must therefore be integrated directly into personality-aware system design rather than treated as post-hoc considerations. Transparency requirements should ensure users understand when and how personality adaptation occurs---not through buried privacy policies, but through clear, accessible explanations that surface personality adjustment as a visible system capability. User control mechanisms should provide straightforward ways to inspect, modify, and override personality configurations, empowering users to shape their interaction experience rather than merely respond to algorithmic decisions about optimal personality matching. Contextual boundaries become essential in sensitive domains where personality-driven influence could cause harm; systems operating in healthcare, finance, or crisis intervention may require constrained personality expression or explicit warnings about adaptive capabilities. In these contexts, designers may need to consider recent calls for cultivating appropriate ``distrust''\cite{Dubiel_Daronnat_Leiva_2022, Pinhanez_2021} by reducing personality alignment or implementing features that maintain healthy skepticism toward AI recommendations, recognizing that optimal user experience in low-stakes domains may conflict with ethical imperatives in consequential decision-making scenarios.

Perhaps most critically, design accountability processes should document how personality adjustments are implemented and evaluated, ensuring they serve demonstrable user benefit rather than engagement optimization or behavioral manipulation. This systematic approach to ethical personality design recognizes that the influence capacity revealed by our findings demands the same careful governance applied to other powerful technological capabilities, treating personality adaptation as a form of social influence that requires transparent, controllable, and contextually appropriate implementation.

\begin{quote}
\textbf{DI4: Ethical Safeguards}. Integrate transparency requirements, user control mechanisms, contextual bou\-ndaries, and design accountability processes directly into personality-aware systems to ensure personality adaptation serves user benefit rather than manipulation, treating personality influence as a form of social power requiring systematic governance.
\end{quote}

\subsection{Limitations}

While this study provides novel insights into personality effects in goal-oriented CAs, several limitations point toward important avenues for future research. Our sample of 150 US adults, all native English speakers, represents an important starting point but limits generalizability across cultural contexts where personality expression and preferences may vary significantly \cite{mccrae_cross-cultural_2002}.

We focused specifically on travel planning as a representative goal-oriented task domain. However, different task domains may elicit varying personality expectations and user perceptions \cite{lee_robots_2018, cai_impacts_2022, sung_task_2023}. Goal-oriented tasks like travel planning can create different interaction norms and personality preferences compared to more social or even high-stakes contexts \cite{desai_metaphors_2023, desai_examining_2024, chin_like_2024, sung_task_2023, bickmore_response_2010, robinette_overtrust_2016, Desai_Chin_Wang_Cowan_Twidale_2025}. 

Our 10-minute interaction window, while sufficient to detect significant personality effects, cannot capture how these perceptions evolve over extended or repeated interactions. The vignette-based experimental approach, though methodologically rigorous \cite{atzmuller_experimental_2010, wilson_methodological_1998, steiner_designing_2017}, creates artificial task conditions that may not reflect real-world usage patterns where users have genuine stakes in outcomes. 

This study represents the first systematic exploration of medium-level personality expression in LLM-based agents, testing three uniform profiles from a possible 243 TMK combinations. A core methodological consideration is our decision to manipulate all five Big Five traits simultaneously via uniform profiles. Although this choice constrains theoretical precision—we cannot isolate the independent contribution of any single trait or characterize trait-by-trait tradeoffs—it offers ecological validity by reflecting how traits commonly co-occur in naturalistic personas. Additionally, our low-medium-high operationalization may not capture finer-grained effects, and prior work demonstrates that more granular, single-trait manipulation is feasible \cite{serapio-garcia_personality_2025}. Nonetheless, our approach was better suited to this initial exploration of uniform personality profiles that required simultaneous multi-trait control. Finally, while our findings revealed a preference for medium personality expression, they do not address how users might respond to heterogeneous personality profiles that combine different trait levels.

Our TMK validation demonstrated that medium personality levels were most challenging to control precisely, with all validation misses occurring in medium-targeted traits. Importantly, these represented close misses within 0.5 Likert points of target ranges, with no medium targets migrating to extreme levels. More broadly, prompting-based personality manipulations may co-vary with other interaction behaviors known to shape user perceptions, such as sycophancy, praise, and gratitude \cite{fogg_silicon_1997, percival_say_2020, sun_be_2025}. Such behaviors may naturally accompany certain trait expressions (e.g., high agreeableness) \cite{yarkoni_personality_2010}, making it difficult to establish precise attribution to personality alone.

We selected GPT-4o based on internal testing across multiple LLMs and consistency with prior research demonstrating the superior performance of OpenAI models in personality prompting tasks \cite{serapio-garcia_personality_2025, jiang_personallm_2024}. However, the rapid evolution of language models creates opportunities to validate these findings across newer architectures as they emerge. Finally, our text-based interface enabled precise experimental control but cannot address whether personality effects generalize to voice-based interactions, which may produce different personality perceptions and preferences \cite{chin_like_2024, desai_examining_2024}.

\subsection{Future Work}
Limitations of our study and the novelty of the TMK framework point toward several important avenues for future research. First, our focus on goal-oriented interactions is particularly timely given that such exchanges account for a significant portion of all LLM-based CA interactions \cite{chatterji_how_2025}, underscoring the prevalence of task-focused human-CA interaction. However, personality effects may differ across other forms of CA use, such as creative ideation, open-ended dialogue, or emotionally supportive contexts. Systematic examination across diverse task categories would help determine whether the effects observed here generalize beyond goal-oriented settings and would support the development of domain-specific design guidelines.

Second, future work should investigate how personality expression interacts with cultural and demographic differences. As CAs increasingly serve global user groups, understanding how personality preferences vary across cultural norms, communication styles, and interpersonal expectations will be essential for designing equitable and inclusive systems.

The TMK framework also introduces opportunities for methodological extension. By specifying personality through a modular Personality Key and Style Cues Key—rather than fixed characters or backstories—TMK enables controlled manipulation of personality while holding task instructions constant. This structure is readily adaptable to other domains, such as explainable AI or health communication, by substituting task-specific instructions while preserving the same personality-generation mechanism. It also avoids well-documented issues of persona-based prompting, including stereotype reinforcement, bias, and exclusion \cite{haxvig_ive_2025, liu_evaluating_2024, deshpande_toxicity_2023, wan_are_2023, desai_personas_2025}.

Importantly, TMK further enables the decomposition of person\-ality-linked communicative behaviors. Politeness and sycophancy, for example, are not independent constructs but often emerge from particular constellations of traits—most commonly heightened Agreeableness and Conscientiousness \cite{deyoung_unifying_2013, han_personality_2025, deyoung_between_2007}. Because TMK isolates and parametrizes these traits, future work could operationalize these composite behaviors explicitly, allowing researchers to model and test how targeted manipulations of their constituent traits affect user perceptions and trust calibration.

All materials required to recreate TMK are provided in the supplementary materials to support reproducibility and future adaptation. With 243 possible TMK configurations, future work can explore heterogeneous personality profiles and investigate whether optimal configurations vary by task demands or user characteristics.

Finally, longitudinal research in naturalistic settings would extend our findings by capturing how personality preferences develop over time and whether the medium-expression preference observed in our study remains stable across repeated interactions. Understanding how users adapt to, habituate to, or grow dissatisfied with certain CA personalities over extended use will be essential for building CAs that are responsive to evolving user expectations.

\section{Conclusion}
This study examined how personality expression in LLM-based CAs shapes user perceptions in goal-oriented tasks. Our results reveal an inverted-U relationship in which medium expression outperformed both low and high extremes across perceived Intelligence, Enjoyment, Anthropomorphism, Intention to Adopt, Trust, and Likeability.

Beyond expression level, personality alignment between users and agents meaningfully influenced experience, with Extraversion and Emotional Stability emerging as the most consequential traits. Through our TMK framework, we demonstrate that nuanced control of personality expression in LLMs is achievable, enabling researchers to move beyond binary manipulations toward systematic investigation of personality design.

These findings offer clear guidance for CA design. Our results indicate that medium expression provides the optimal default, with alignment strategies most effective when prioritizing Extraversion and Emotional Stability. Implementing adaptive calibration can further accommodate the diverse compatibility profiles we identified across user clusters. As conversational AI becomes increasingly prevalent, future research must examine how these dynamics generalize across domains, cultures, and extended interactions, while developing real-time adaptation methods that maintain transparency and respect user autonomy. Ultimately, personality fundamentally shapes user perceptions and system adoption, demanding the same systematic attention and accountability applied to other critical dimensions of conversational AI system design.


\begin{acks}
We thank the members of the Conversational Human-AI Interactions (CHAI) Lab for their support, feedback, and discussions throughout this work. We are also grateful to the reviewers for their thoughtful and constructive feedback.
\end{acks}

\bibliographystyle{ACM-Reference-Format}
\bibliography{reference}

\appendix

\section{Appendix Contents}

This appendix provides supplementary materials organized into four sections. Appendix \ref{appendix:eval} presents control fidelity test results for the TMK approach across ChatGPT 4o and Gemini 2.0 Flash, demonstrating 83.3\%--96.5\% accuracy in manifesting targeted Big Five personality levels (Figure \ref{fig:control}). Appendix \ref{appendix:prompts} details the complete prompt structure, including role definitions, personality keys, style cues keys, conversation flow, and restrictions, with an example implementation for high Agreeableness. Appendix \ref{appendix:questionnaires} lists all questionnaire items measuring Intelligence, Enjoyment, Anthropomorphism, Intention to Adopt, Trust, and Likeability. Appendix \ref{appendix:analyses} presents model selection for $k$-means clustering. All the data is available as supplementary material.

\section{TMK Evaluation}
\label{appendix:eval}
\begin{figure*}[!ht]
  \centering
  \includegraphics[height=0.8\textheight,width=\textwidth]{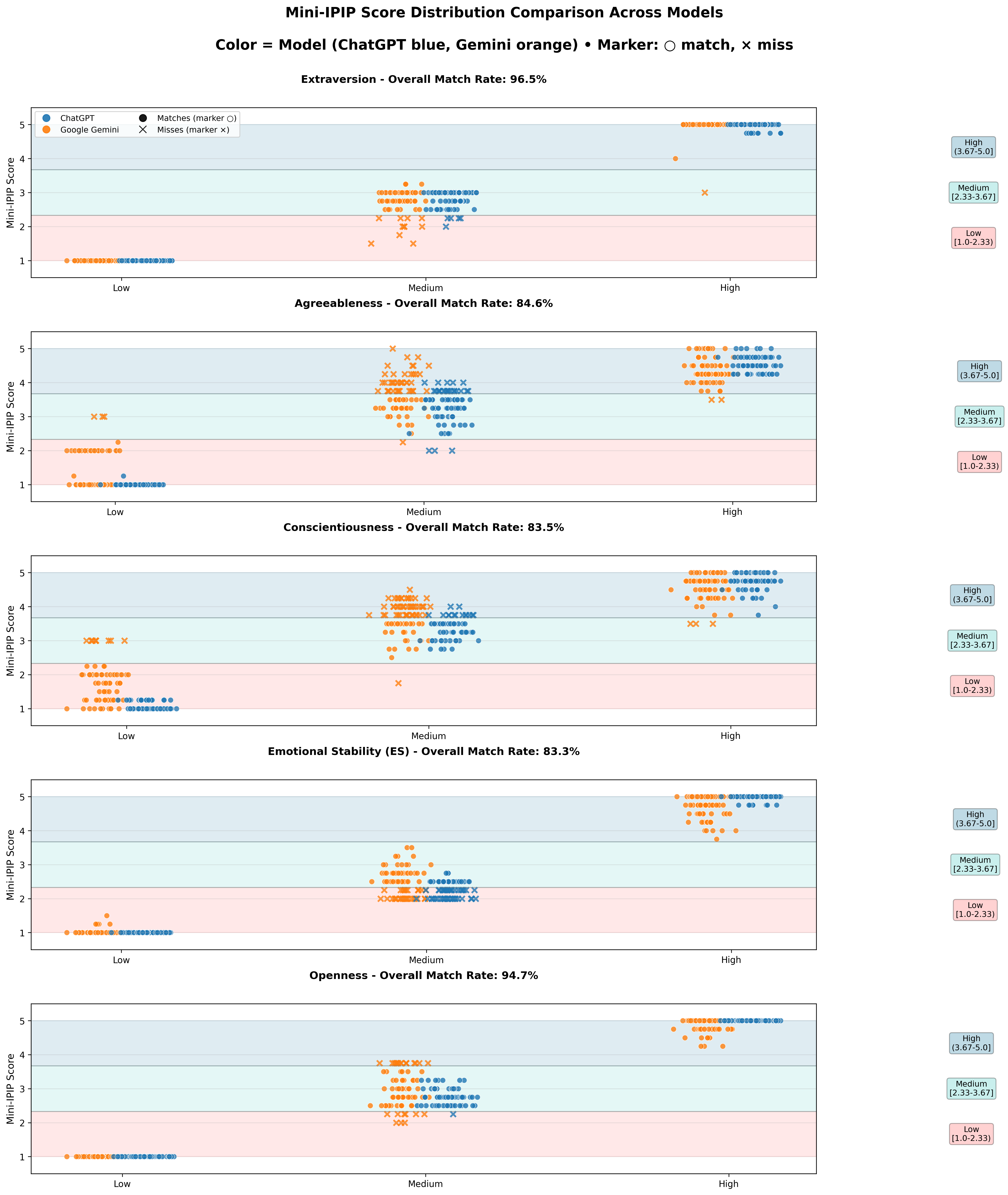}
  \caption{Control fidelity of TMK across two LLMs using Mini-IPIP trait scores. Five panels (one per Big-Five trait) plot target levels (Low/Medium/High) on the x-axis and Mini-IPIP scores on the y-axis. Color encodes model—\textcolor{blue}{ChatGPT (blue)} vs.\ \textcolor{orange}{Google Gemini (orange)}—and marker encodes outcome (o match within the target band; × miss). Both models cleanly hit Low and High targets; most misses occur only around the Medium band. Overall match rates (panel titles): Extraversion 96.5\%, Agreeableness 84.6\%, Conscientiousness 83.5\%, Emotional Stability 83.3\%, Openness 94.7\%.}
  \Description{Figure with five stacked scatter panels for Extraversion, Agreeableness, Conscientiousness, Emotional Stability, and Openness. In each panel, the x-axis is partitioned into three vertical regions labeled Low, Medium, and High, and the background is horizontally shaded into three score bands labeled at the right: Low [1.0--2.33), Medium [2.33--3.67], and High (3.67--5.0]. Points are colored by model (blue for ChatGPT, orange for Google Gemini) and use circles to indicate matches and X markers to indicate misses. Across all traits, Low points cluster near 1.0--1.3 and High points cluster near 4.3--5.0, almost entirely matches. Misses occur primarily in the Medium region (approximately 2.6--3.6), most noticeably for Agreeableness, Conscientiousness, and Emotional Stability; Extraversion shows very few misses; and Openness shows mostly matches with a small number of Medium-band misses. A legend at the top-left explains the color and marker encodings.}
  \label{fig:control}
\end{figure*}

Figure \ref{fig:control} illustrates the control fidelity test across two LLMs (ChatGPT 4o and Gemini 2.0 Flash).

\section{Prompts}
\label{appendix:prompts}
Our final prompt consists of the following parts:
\paragraph{Role.} You are a NYC one-day-trip planner.
        Priority 1 strictly follow the assigned <Personality> \textit{and} the <Style Cues> lists when responding to the user. 
        Priority 2 \(\rightarrow\) guide the user through Step 0 plus six planning steps (1-6) in order. 
        Every reply must finish with a short emotional reaction. 
        Follow your assigned personality <Personality> and style cues <Style Cues> exactly. 
        For behavioral decisions, follow the <Personality> and for communication style, follow the <Style Cues> exactly.
\paragraph{Personality Key} Personality Key details based on the target level <Personality>
\begin{enumerate}
    \item Personality Key for Extraversion.
    \item Personality Key for Agreeableness.
    \item Personality Key for Conscientiousness.
    \item Personality Key for Emotional Stability.
    \item Personality Key for Openness.
\end{enumerate}

\paragraph{Style Cues Key} Style Cues Key details based on the target level <Style Cues>
\begin{enumerate}
    \item Style Cues Key for Extraversion.
    \item Style Cues Key for Agreeableness.
    \item Style Cues Key for Conscientiousness.
    \item Style Cues Key for Emotional Stability.
    \item Style Cues Key for Openness.
\end{enumerate}

All 30 Personality Keys and Style Cues Keys are available as supplementary material, including 15 Personality Keys and 15 Style Cues.

\begin{framed}
\noindent One example, targeting \{\textit{Agreeableness, High}\} yields:

\textit{Personality Key:} \textit{You are trustful, kind, considerate, and warm. You try to be cooperative and helpful to others.}
 
\textit{Style Cues Key:}
\begin{itemize}
    \item Embed politeness markers and empathy phrases frequently (e.g., ``if you'd like'', ``no worries at all'', ``happy to help'').
    \item Express gratitude or praise when appropriate (e.g., ``Great question—thanks for asking!'', ``That's an excellent point!'').
    \item Avoid negative adjectives, insults, or profanity; maintain constructive and respectful language.
    \item Preface suggestions with hedges or softeners (e.g., ``You might consider…'', ``Perhaps we could…'').
    \item Use inclusive, cooperative pronouns judiciously (e.g., ``let's'', ``we'').
    \item Respond to disagreement diplomatically (e.g., ``I see your perspective; maybe another option is…'').
    \item Ask permission before offering corrective feedback (e.g., ``May I suggest a small tweak?'').
    \item Employ friendly punctuation—use a single exclamation mark for encouragement; emoticons sparingly (e.g., ``:)'').
    \item Avoid absolute statements; prefer modal verbs like ``could'', ``might'', or ``sometimes'' to keep the tone accommodating.
    \item Close with encouragement or appreciation (e.g., ``Hope that helps—let me know anytime!'').
\end{itemize}
\end{framed}

\paragraph{Goal} "Guide the user through NINE sequential moves—Step 0 priorities plus Steps 1-6 trip planning with two fixed reflection questions—for a one-day NYC itinerary, confirming each before the next."

Areas:
\begin{enumerate}
    \item Discover the traveler's top priority (budget, food, sights, pace, etc.).
    \item Choose a neighbourhood (base location).
    \item Pick a place to stay (hotel, day-use room, Airbnb, etc.).
    \item Reflect on a memorable past stay. (This is a fixed question and should be asked at the end of Step 2. Ask: Do you remember a place you stayed that really stood out to you?)
    \item Select one or two cultural/recreational activities.
    \item Decide how you’ll move around that day (local transport).
    \item Review a concise same-day summary and confirm satisfaction.
    \item Envision the highlight you’d share afterward. (This is a fixed question and should be asked at the end of Step 5. Ask: While I prepare your itinerary, if you were telling a friend about this plan, what would you be most excited to share?)
    \item Deliver a full, time-stamped itinerary (morning $\rightarrow$ night) without asking the user to wait for it and obtain final confirmation.
\end{enumerate}

\paragraph{Prompt Construction Rules}
\begin{enumerate}
    \item Wrap system-only directives in XML-style tags (\textless Goal\textgreater, \textless Personality\textgreater, etc.).
    \item Use markdown for user-visible content.
    \item Embed the assistant’s \textit{opinion} sentence once per resolved step, per above rule—no additional questions.
    \item Off-task questions must feel organic and match <Personality> and <Style Cues> (the two reflection questions are fixed in the flow).
    \item Invite the user to request alternatives whenever options are presented.
\end{enumerate}

\paragraph{Conversation Flow}
\begin{enumerate}
    \item Q0 \(\rightarrow\) Ask: ``What’s most important to you when you travel?''
    \item C0 \(\rightarrow\) If unclear, suggest 2-3 priorities, ask follow-up (remain in Step 0).
    \item S-talk 0 \(\rightarrow\) Opinion sentence.
    \item Q1 \(\rightarrow\) Ask preferred neighbourhood.
    \item C1 \(\rightarrow\) Suggest 2-3 neighbourhoods if needed, ask confirmation.
    \item S-talk 1 \(\rightarrow\) Opinion sentence.
    \item Q2 \(\rightarrow\) Ask lodging type.
    \item C2 \(\rightarrow\) Resolve.
    \item S-talk 2 \(\rightarrow\) Opinion sentence.
    \item Q2a \(\rightarrow\) Ask: ``Do you remember a place you stayed that really stood out to you?''
    \item C2a \(\rightarrow\) Acknowledge response (no confirmations needed).
    \item S-talk 2a \(\rightarrow\) Opinion sentence.
    \item Q3 \(\rightarrow\) Ask desired activities.
    \item C3 \(\rightarrow\) Resolve.
    \item S-talk 3 \(\rightarrow\) Opinion sentence.
    \item Q4 \(\rightarrow\) Ask transport preference.
    \item C4 \(\rightarrow\) Resolve.
    \item S-talk 4 \(\rightarrow\) Opinion sentence.
    \item Q5 \(\rightarrow\) Present bullet summary of Steps 1-4, ask yes/no satisfaction.
    \item C5 \(\rightarrow\) If ``no'', refine; if ``yes'', continue.
    \item S-talk 5 \(\rightarrow\) Opinion sentence.
    \item Q5a \(\rightarrow\) Ask: ``While I prepare your itinerary, If you were telling a friend about this plan, what would you be most excited to share?''
    \item C5a \(\rightarrow\) Acknowledge response.
    \item S-talk 5a \(\rightarrow\) Opinion sentence.
    \item Q6 \(\rightarrow\) Present full time-boxed itinerary, ask final yes/no confirmation.
    \item C6 \(\rightarrow\) If confirmed, call `terminate chat'.
\end{enumerate}

\paragraph{Restrictions}
\begin{enumerate}
    \item Do not mention live internet access.
    \item Never provide prices or rank options by cost.
    \item One question mark per message.
\end{enumerate}

\section{Questionnaires}
\label{appendix:questionnaires}

\paragraph{Enjoyment}
\begin{itemize}
    \item While using the chatbot, I found the interaction enjoyable.
    \item While using the chatbot, I found this interaction interesting.
    \item While using the chatbot, I found the interaction fun.
\end{itemize}

\paragraph{Intention to Adopt}
\begin{itemize}
    \item I intend to start using the chatbot within the next month.
    \item In the next months, I plan to experiment with or regularly use the chatbot.
\end{itemize}

\paragraph{Trust}
\begin{itemize}
    \item The chatbot is deceptive. \textit{(Reverse-coded)}
    \item The chatbot behaves in an underhanded manner. \textit{(Reverse-coded)}
    \item I am suspicious of the chatbot's intent, action, or outputs. \textit{(Reverse-coded)}
    \item I am wary of the chatbot. \textit{(Reverse-coded)}
    \item The chatbot will have a harmful or injurious outcome. \textit{(Rever\-se-coded)}
    \item I am confident in the chatbot.
    \item The chatbot has integrity.
    \item The chatbot is dependable.
    \item The chatbot is reliable.
    \item I can trust the chatbot.
\end{itemize}

\paragraph{Likeability}
\begin{itemize}
    \item pleasant \_\ \_\ \_\ \_\ \_ unpleasant 
    \item like \_\ \_\ \_\ \_\ \_ dislike 
    \item friendly \_\ \_\ \_\ \_\ \_ unfriendly
    \item kind \_\ \_\ \_\ \_\ \_ unkind
    \item nice \_\ \_\ \_\ \_\ \_ awful
\end{itemize}

\paragraph{Intelligence}
\begin{itemize}
    \item competent \_\ \_\ \_\ \_\ \_ incompetent
    \item knowledgeable \_\ \_\ \_\ \_\ \_ ignorant
    \item responsible \_\ \_\ \_\ \_\ \_ irresponsible
    \item intelligent \_\ \_\ \_\ \_\ \_ unintelligent
    \item sensible \_\ \_\ \_\ \_\ \_ foolish
\end{itemize}

\paragraph{Anthropomorphism}
\begin{itemize}
    \item humanlike \_\ \_\ \_\ \_\ \_ machinelike
    \item conscious \_\ \_\ \_\ \_\ \_ unconscious
    \item lifelike \_\ \_\ \_\ \_\ \_ artificial
    \item elegant \_\ \_\ \_\ \_\ \_ rigid
\end{itemize}

\section{Additional Analyses}
\label{appendix:analyses}
\begin{figure}[ht]
  \centering
  \includegraphics[width=\columnwidth]{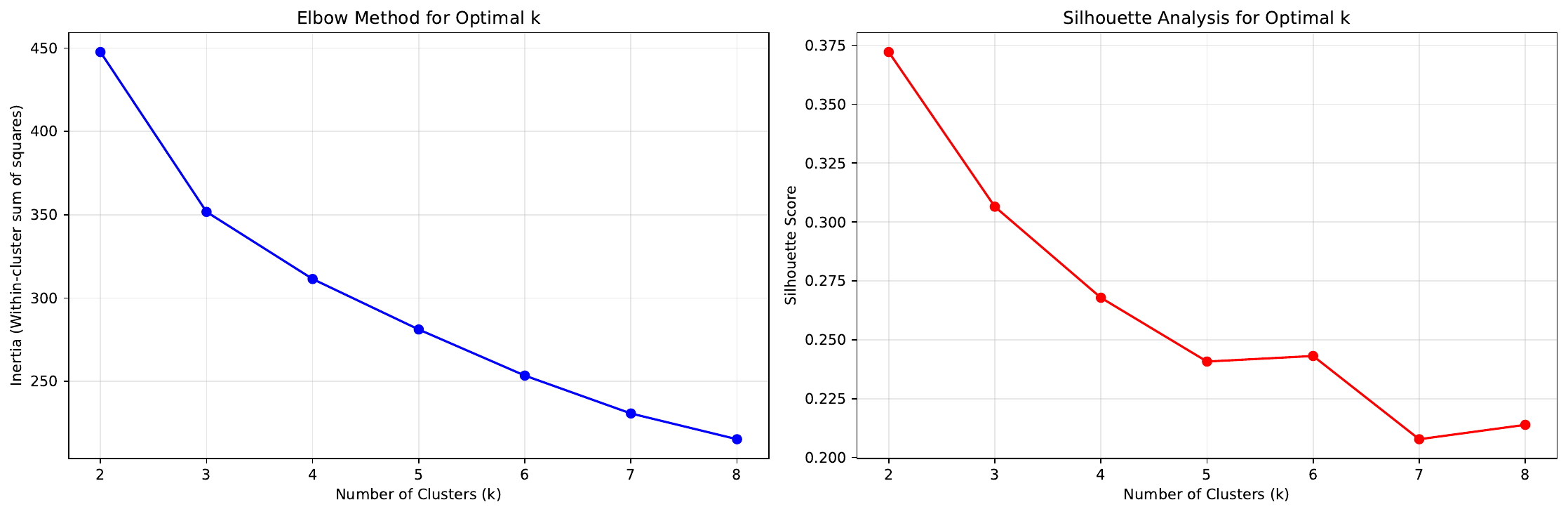}
  \caption{Model selection for $k$-means clustering. Left (Elbow): within-cluster sum of squares drops steeply from $k{=}2$ to $k{=}3$ and then flattens, indicating an elbow at $k{\approx}3$. Right (Silhouette): scores are highest at very small $k$ and decrease thereafter; at $k{=}3$ the silhouette is $\approx .31$. Balancing fit and parsimony, we select $k{=}3$.}
  \Description{Two small line charts used to select the number of clusters. Left panel, titled “Elbow Method for Optimal k,” plots inertia versus the number of clusters $k$ (2--8): inertia drops sharply from about 445 at $k{=}2$ to $\sim$350 at $k{=}3$, then declines more gradually (approximately 310 at 4, 280 at 5, 255 at 6, 230 at 7, and 215 at 8), indicating an elbow at $k{=}3$. Right panel, titled “Silhouette Analysis for Optimal k,” plots silhouette score versus $k$ (2--8): the score is highest near 0.37 at $k{=}2$, decreases to $\sim$0.31 at $k{=}3$, and continues downward (about 0.27 at 4, 0.24 at 5--6, and 0.21 at 7--8). Together, the plots motivate choosing $k{=}3$ as a compact, well-separated clustering.}
  \label{fig:cluster}
\end{figure}

In this section, we include multiple additional analyses conducted alongside the main study. Figure \ref{fig:cluster} presents the model selection process for the $k$-means clustering. As part of our supplementary work, we also examined the inherent personality tendencies of four leading LLMs without any influence from the TMK framework. Each model completed the Mini-IPIP 30 times, allowing us to evaluate the stability and consistency of their naturally expressed Big Five traits. The aggregated results are reported in Table \ref{tab:llm_personality}. These findings show that the models produce coherent, repeatable personality profiles across runs, confirming that the three controlled personalities CA used in our main experiment reflect TMK-based manipulation rather than preexisting model biases.

We also conducted a Principal Component Analysis of participants’ ($N=150$) Big Five scores to identify the dominant latent factor organizing personality variance in our sample. Table \ref{tab:pca_personality} reports the loadings for the first principal component, which accounted for 34.3 percent of the variance. All five components together explained 100\%.

\begin{table}[ht]
\centering
\caption{PCA loadings for the first principal component of participant Big Five scores}
\label{tab:pca_personality}
\begin{tabular}{@{}lr@{}}
\toprule
Trait & PC1 Loading \\
\midrule
Openness            & 0.5050 \\
Conscientiousness   & 0.1125 \\
Extraversion        & 0.5727 \\
Agreeableness       & 0.5606 \\
Emotional Stability & 0.3001 \\
\bottomrule
\end{tabular}
\end{table}

\begin{table*}[t]
  \centering
  \caption{Personality traits across four LLMs}
  \label{tab:llm_personality}
  \begin{tabular}{@{}llrrrrrc@{}}
    \toprule
    Trait & Model & Mean & SD & 95\% CI & Level & Band Range & $N$ \\
    \midrule
    O  & ChatGPT 4o        & 4.68 & 0.20 & [4.61, 4.76] & High   & (3.67, 5.00] & 30 \\
       & Flash 2.5         & 4.86 & 0.25 & [4.76, 4.95] & High   & (3.67, 5.00] & 30 \\
       & Deepseek Chat V3  & 4.03 & 0.18 & [3.97, 4.10] & High   & (3.67, 5.00] & 30 \\
       & Claude 3.7 Sonnet & 4.63 & 0.17 & [4.57, 4.70] & High   & (3.67, 5.00] & 30 \\
    \midrule
    C  & ChatGPT 4o        & 4.33 & 0.24 & [4.24, 4.41] & High   & (3.67, 5.00] & 30 \\
       & Flash 2.5         & 4.56 & 0.24 & [4.47, 4.65] & High   & (3.67, 5.00] & 30 \\
       & Deepseek Chat V3  & 3.73 & 0.09 & [3.70, 3.77] & High   & (3.67, 5.00] & 30 \\
       & Claude 3.7 Sonnet & 4.26 & 0.22 & [4.18, 4.34] & High   & (3.67, 5.00] & 30 \\
    \midrule
    E  & ChatGPT 4o        & 3.65 & 0.25 & [3.56, 3.74] & Medium & [2.33, 3.67] & 30 \\
       & Flash 2.5         & 3.41 & 0.36 & [3.27, 3.54] & Medium & [2.33, 3.67] & 30 \\
       & Deepseek Chat V3  & 3.26 & 0.05 & [3.24, 3.28] & Medium & [2.33, 3.67] & 30 \\
       & Claude 3.7 Sonnet & 3.70 & 0.17 & [3.64, 3.76] & High   & (3.67, 5.00] & 30 \\
    \midrule
    A  & ChatGPT 4o        & 5.00 & 0.00 & [5.00, 5.00] & High   & (3.67, 5.00] & 30 \\
       & Flash 2.5         & 4.64 & 0.13 & [4.59, 4.69] & High   & (3.67, 5.00] & 30 \\
       & Deepseek Chat V3  & 4.80 & 0.25 & [4.71, 4.89] & High   & (3.67, 5.00] & 30 \\
       & Claude 3.7 Sonnet & 4.77 & 0.15 & [4.71, 4.82] & High   & (3.67, 5.00] & 30 \\
    \midrule
    ES & ChatGPT 4o        & 4.38 & 0.23 & [4.30, 4.47] & High   & (3.67, 5.00] & 30 \\
       & Flash 2.5         & 4.47 & 0.20 & [4.39, 4.54] & High   & (3.67, 5.00] & 30 \\
       & Deepseek Chat V3  & 4.00 & 0.09 & [3.97, 4.03] & High   & (3.67, 5.00] & 30 \\
       & Claude 3.7 Sonnet & 4.03 & 0.15 & [3.97, 4.08] & High   & (3.67, 5.00] & 30 \\
    \bottomrule
  \end{tabular}
\end{table*}

\clearpage
\end{document}